\documentclass[a4paper,11pt]{article}

\title{An extensive survey of the estimation of
uncertainties from missing higher orders in perturbative calculations}
\author{ E. Bagnaschi$^{1,2,3}$, M. Cacciari$^{1,2,3}$, A. Guffanti$^{4}$ and L. Jenniches$^{4}$
\vspace{0.7cm}
\\
{\sl  \small $^1$Universit\'e Paris Diderot, Paris, France}\\[2pt]
{\sl  \small $^2$Sorbonne Universit\'es, UPMC Univ Paris 06, UMR 7589, LPTHE, F-75005, 
Paris, France}\\[2pt]
{\sl  \small $^3$CNRS, UMR 7589, LPTHE, F-75005, Paris, France}\\[2pt]
{\sl  \small $^4$Niels Bohr International Academy and Discovery Center,}\\
{\sl  \small Niels Bohr Institute, University of Copenhagen,}\\
{\sl  \small Blegdamsvej 17, DK-2100 Copenhagen, Denmark}\\[2pt]
}
\date{}

\usepackage[english]{babel}
\usepackage[utf8]{inputenc}
\usepackage[T1]{fontenc}
\usepackage{times}
\usepackage{color}
\usepackage{hyperref}
\usepackage{graphicx}
\usepackage{fullpage}
\usepackage{amssymb,amsmath}
\usepackage{pdfpages}
\usepackage{float}
\usepackage{morefloats}
\usepackage{mathtools}
\usepackage{caption}
\usepackage{bm}
\usepackage{booktabs}
\usepackage{subfigure}
\usepackage{comment}
\usepackage{pdflscape}

\numberwithin{equation}{section}
\numberwithin{figure}{section}
\newcommand{\be}{\begin{equation}}
\newcommand{\ee}{\end{equation}}
\newcommand{\ba}{\begin{eqnarray}}
\newcommand{\ea}{\end{eqnarray}}

\newcommand{\chbar}{$\overline{\mathrm{CH}}$}

\newcommand{\as}{\alpha_s}

\newcommand{\ra}[1]{\renewcommand{\arraystretch}{#1}}

\begin{document}

\maketitle
\vspace{-10.5cm}
\begin{flushright}
        \begin{minipage}{3.0cm}
          LPN14-115
        \end{minipage}
\end{flushright}
\vspace{10cm}
\begin{abstract}
We consider two approaches to estimate and characterise the theoretical uncertainties 
stemming from the missing higher orders in perturbative calculations in Quantum Chromodynamics: 
the traditional one based on renormalisation and factorisation scale variation, and the 
Bayesian framework proposed by Cacciari and Houdeau.
We estimate uncertainties with these two methods for a comprehensive set of more than thirty different observables 
computed in perturbative Quantum Chromodynamics, and we discuss their performance in properly estimating the 
size of the higher order terms that are known.
We find that scale variation with the conventional choice of varying scales within a factor 
of two of a central scale gives uncertainty intervals that tend to be somewhat too small to be 
interpretable as 68\% confidence-level-heuristic  ones.
We propose a modified version of the Bayesian approach of Cacciari and Houdeau which
performs well for non-hadronic 
observables and, after an appropriate choice of the relevant expansion parameter for the perturbative series, for hadronic 
ones too.
\end{abstract}
\thispagestyle{empty}

\newpage
\tableofcontents
\clearpage

\section{Introduction}
\label{sec:intro}

Precision phenomenology of the kind aimed for by the Large Hadron Collider (LHC) physics 
program requires accurate and reliable theoretical predictions to be compared to an ever increasing
range of high precision experimental measurements. Once theoretical and experimental uncertainties
become of comparable size, it is crucial to be able to characterise quantitatively the relevance of missing 
higher order terms in perturbative calculations.

In Quantum Chromodynamics (QCD), which we take as a model here given its 
central role in LHC physics, theoretical uncertainties stemming from missing higher orders 
in the perturbative series are usually estimated by varying the unphysical renormalisation 
and factorisation scales that appear in the cross-sections and decay rates calculations. This approach has served 
the QCD community well for more than thirty years, and can still be regarded as the most 
effective way to quickly estimate the missing higher order uncertainties (MHOUs). 
It suffers, however, from some drawbacks. Chiefly among them the fact that its uncertainty 
intervals cannot be characterised in a statistically meaningful way and 
therefore cannot be combined easily with, e.g., likelihood profiles for other 
uncertainties, for instance of experimental origin.

One of us (MC) and N.~Houdeau tried in~\cite{Cacciari:2011ze} to overcome this limitation 
by proposing to estimate MHOUs in a Bayesian context, so as to obtain a statistically 
meaningful posterior distribution for the probability density profile of the uncertainty interval. 
The Cacciari-Houdeau approach led to a model (henceforth CH) that relies on simple priors 
that, at their core, partly mimic assumptions that are anyway implicitly made when one employs 
the scale-variation method. We refer to~\cite{Cacciari:2011ze} for a more detailed description 
of the CH approach and its underlying Bayesian character, and e.g. to \cite{Ball:2011us,
Goria:2011wa,Forte:2013mda} for some examples of applications of its results. In a context of estimation of MHOUs, we also point out the different but possibly complementary approach of~\cite{David:2013gaa} that focuses on a mathematically motivated approximate completion of a perturbative series.

The purpose of this paper is twofold. On the one hand, we revisit the Bayesian CH model, and 
propose a modified version (which we will denote \chbar) which will trade some of 
the simplicity of the original CH model for a better adaptability to a broader class of 
observables, namely those related to processes with hadrons in the initial state. On the other 
hand, we  study the results of both the scale-variation and the \chbar\ model on a large 
number of perturbatively calculated observables, so as to be able to assess their performance 
in a (frequentist) statistically meaningful way. For the scale-variation approach, this means 
that we can attempt to characterise {\sl a posteriori} its uncertainty intervals in terms 
of some confidence level that they correctly describe the MHOUs. For the \chbar\ model, this 
study allows us to either assess  whether the {\sl Degree of Belief} (DoB) associated to the 
uncertainty intervals is correct or, where needed, to estimate the appropriate expansion parameter of the perturbative series  
that ensures that this be the case.

The paper is structured as follows. Section~\ref{sec:desc} reviews the scale-variation
approach and the Bayesian method introduced in \cite{Cacciari:2011ze}, and describes the 
modifications to the CH model that lead to the formulation of the \chbar\ approach used in this 
paper. 
Section~\ref{sec:global} describes the methodology that we have followed in our study of the 
performances of the scale-variation and the \chbar\ approaches, introduces the list of calculated
observables used in the survey, and presents our results. 
Section~\ref{sec:benchmark} compares the results of the scale-variation and the \chbar\ 
method for the determination of MHOUs for some benchmark processes that we consider either 
particularly relevant for LHC phenomenology or simply quite iconic, namely $e^+e^-\to$ hadrons, 
Higgs decay to two gluons and to two photons, $W$ and $Z$ production in $pp$ collisions, $pp\to t\bar t$ 
and Higgs production in proton-proton collisions. 
A concluding section follows, while a few appendices collect technical details and the 
numerical values of the perturbative coefficients of the observables used in the survey 
and the benchmarking.

\section{Estimations of theoretical uncertainties}
\label{sec:desc}

In this Section we introduce and describe two different approaches to the estimation of the uncertainty stemming from the missing higher orders of a perturbatively  calculated observable:

\begin{itemize}
\item the scale-variation approach, which involves varying the unphysical renormalisation and 
factorisation scales that appear in higher order perturbative calculations within a given range 
around a chosen central value;
\item the Bayesian approach introduced by Cacciari and Houdeau in \cite{Cacciari:2011ze}, 
with its modification discussed below.
\end{itemize}
In the following we review how these two approaches work, and also set the appropriate notations.

\subsection{Uncertainty estimation by scale variation}\label{sec:th-sv}

The truncated perturbative expansion of an arbitrary observable $O$ calculated up to a 
fixed order $k$ as a power series expansion in $\alpha_s$, 
\begin{align}
  O_k(Q,\mu) = \sum\limits_{n=l}^{k}\alpha_s^n(\mu) c_n(Q,\mu) \,,
\end{align}
contains a residual, higher-order dependence on the renormalisation and/or factorisation scales, here collectively denoted by $\mu$.

The standard approaches to estimate the MHOUs are all based on the idea of varying the scale(s) 
$\mu$ in an interval $[Q/r,rQ]$, where $r$ is an arbitrary factor often chosen to be equal 
to 2, and $Q$ is a typical hard scale of the process.
The values of the observable obtained at different scales are then used to derive an 
uncertainty interval. Different recipes can be used to implement this prescription.
Writing this interval as $[O_k^-,O_k^+]$ around $O_k$ (not necessarily centred around it), 
the most common choices are:
\begin{enumerate}
\item
  \begin{align}
    O_k^{-} &= \text{min} \{O_k(Q,Q/r),O_k(Q,rQ), O_k(Q,Q)\}\,, \nonumber \\
    O_k^{+} &= \text{max} \{O_k(Q,Q/r),O_k(Q,rQ), O_k(Q,Q) \}\,.
    \label{eq:scale1}
  \end{align}
\item
  \begin{align}
    O_k^{-} = \min_{\mu \in [Q/r,rQ]} \{O_k(Q,\mu)\}\,, \qquad  \qquad O_k^{+} = 
    \max_{\mu \in [Q/r,rQ]} \{O_k(Q,\mu)\}
    \label{eq:scale2}
  \end{align}
\item
  \begin{align}
    O_k^\pm = O_k \pm \frac{\delta_k}{2}\,,
    \label{eq:scale3}
  \end{align}
  where we have defined
  \begin{align}
    \delta_k \equiv | O_k(Q,rQ) - O_k(Q,Q/r) |\,.
  \end{align}
\item Same as eq.~(\ref{eq:scale3}) but with
  \begin{align}
    \delta_k\equiv\max_{\mu \in [Q/r,rQ]}\{O_k(Q,\mu)\}-\min_{\mu \in [Q/r,rQ]}\{O_k(Q,\mu)\}\,.
  \end{align}
\end{enumerate}

Generalisation to the case of two or more scales is straightforward and follows along the 
same lines. The  prescription which is probably most commonly used (see e.g. the QCD review in~\cite{Agashe:2014kda}) , and which we will also use in our study, is 
 an extension of eq.~(\ref{eq:scale1}), i.e. varying both the renormalisation and 
the factorisation scale ($\mu_r$ and $\mu_f$) as shown there, but with the additional constraint 
$1/r \leq \mu_r/\mu_f \leq r$, to avoid the appearance of unnaturally large logarithms.\footnote{To the best of our knowledge, this additional constraint was first adopted in \cite{Cacciari:2003fi}, following a suggestion by Stefano Catani.}

The main problem with the scale-variation approach is that it does not provide a probability distribution 
for the uncertainty interval, which therefore has no statistical meaning. It is also worth 
noting that the common choice $r=2$ is merely a convention, and that the choice of the central 
scale around which to perform the variation is also largely arbitrary. In fact, in some cases 
this central scale is deliberately chosen away from the characteristic scale of the process 
to satisfy other criteria. This is for instance the case for Higgs production in gluon fusion,
where the central scale is often chosen equal to $m_H/2$ to mimic the result obtained when 
performing soft-gluon resummation~\cite{Anastasiou:2005qj}, and because around this value 
the cross section shows reduced sensitivity to the scale choice and an improved convergence 
of the perturbative series~\cite{Anastasiou:2002yz}.

\subsection{The Cacciari-Houdeau Bayesian approach}

The approach of Cacciari and Houdeau~\cite{Cacciari:2011ze} is a Bayesian 
 framework to evaluate MHOUs. It makes assumptions on the behaviour of the 
coefficients of a series of the form

\begin{equation}
  \label{eq:chexp}
  O_k \equiv O_k(Q,Q)=\sum\limits_{n=l}^{k}\alpha_s^n(Q) c_n(Q,Q) \equiv \sum\limits_{n=l}^{k}\alpha_s^n c_n \, ,
\end{equation} 
where the unphysical scales $\mu$ have been set to the central value $Q$, and we have implicitly 
defined $\alpha_s \equiv \alpha_s(Q)$ and $c_n\equiv c_n(Q,Q)$. These assumptions are encoded 
into specific Bayesian priors (detailed in \cite{Cacciari:2011ze}) and in the choice of the expansion parameter, 
taken here to be $\alpha_s$, and allow one to determine an uncertainty density profile (the posterior of the 
model) in the form of a conditional probability density for the remainder of the series\footnote{The use of an upper limit for the summation at infinity in $\Delta_k$ should be considered as merely symbolic, QCD series being asymptotic. In practice, the remainder that we will be dealing with will be limited to the region of apparent convergence of the series, and will be usually approximated by its first term.}, $\Delta_k\equiv\sum_{n=k+1}^\infty\alpha_s^n c_n$, 
given the known perturbative coefficients,  $\{c_l,\dots,c_k\}$.
Assuming that the dominant contribution to the remainder comes from the first unknown order, i.e. 
$\Delta_k \simeq \alpha_s^{k+1} c_{k+1}$, one  can derive~\cite{Cacciari:2011ze} a simple analytic expression for the conditional
density, 
\be
\label{eq:DeltaKnowCkExpression}
f(\Delta_k|c_l,\dots,c_k)\simeq\left(\frac{n_c}{n_c+1}\right)
\frac{1}{2\as^{k+1}\bar c_{(k)}}
\left\{ 
\begin{array}{cc}
  1&\mbox{ if }	|\Delta_k|\leq \as^{k+1}\bar c_{(k)}\\[10pt]
  \left(\frac{\alpha_s^{k+1}\bar{c}_{(k)}}{|\Delta_k|}\right)^{n_c+1} &\mbox{ if } |\Delta_k|>\as^{k+1}\bar c_{(k)}
\end{array}
\right. \, ,
\ee
where $\bar{c}_{(k)}\equiv\max(|c_l|,\cdots,|c_k|)$ and $n_c$ is the number of known perturbative 
coefficients.
From this expression, one can appreciate the characteristics of the posterior distribution 
for this model: a central plateau with power suppressed tails.

The existence of such a probability density distribution for the uncertainty interval represents
the main difference with the scale-variation approach, which only gives an interval without 
a density profile.

Given the conditional density in eq.~(\ref{eq:DeltaKnowCkExpression}) it is possible to compute
the smallest credibility interval for $\Delta_k$ with a degree of belief (DoB) equal to $p\%$ (i.e. such 
that $\Delta_k$ is expected with $p\%$ credibility to be contained within the interval $[-d_k^{(p)}, d_k^{(p)}]$) :
\begin{align}
  d_k^{(p)} & =\left\{
  \begin{array}{l l} \alpha_s^{k+1} \bar{c}_{(k)} \frac{n_c+1}{n_c} p\% & 
  \text{if} \qquad p\% \leq \frac{n_c}{n_c+1} \\\\
  \alpha_s^{k+1} \bar{c}_{(k)} \left[(n_c+1)(1-p\%)\right]^{(-1/n_c)} & \text{if} \qquad 
  p\% > \frac{n_c}{n_c+1}   \\
  \end{array}\right.
\end{align}

\subsection{The modified Cacciari-Houdeau approach \texorpdfstring{($\overline{\mathrm{CH}}$)}{}}
\label{sec:chbarfac}
The CH model described above relies on a  specific form of the perturbative expansion, 
namely eq.~(\ref{eq:chexp}). As a result, its
estimate for the uncertainty is not invariant under a rescaling of the
expansion parameter from $\as$ to $\as/\lambda$. 
While working on this project  we made a number of attempts to reformulate the model in a rescaling-invariant 
way. Ultimately, none of them turned out to be satisfactory, to the extent that each required formulating priors 
much too informative, which shaped excessively the final posterior.
We eventually settled instead on a slightly modified version of the CH model. In this modified model, henceforth 
denoted as \chbar, we rewrite the perturbative expansion of eq.~(\ref{eq:chexp}) in the form
\begin{equation}
  \label{eq:chbarexp}
  O_k=\sum\limits_{n=l}^{k}\frac{\alpha_s^n}{\lambda^n} (n-1)!
  \frac{\lambda^n c_n}{(n-1)!}
  \equiv
  \sum\limits_{n=l}^{k}\left(\frac{\alpha_s}{\lambda}\right)^n (n-1)!\, b_n\, ,
\end{equation}
with
\be
b_n \equiv \frac{\lambda^n c_n}{(n-1)!} \, ,
\ee
and submit the new coefficients $b_n$ to the same priors originally used for the $c_n$ in the CH model. 
This leads to the following expressions for the probability density profile for the remainder function $\Delta_k$
\be
f(\Delta_k|b_l,\dots,b_k)\simeq\left(\frac{n_c}{n_c+1}\right)
\frac{1}{2k!(\as/\lambda)^{k+1}\bar b_{(k)}}\left\{
\begin{array}{cc}
  1    & \mbox{ if }	|\Delta_k|\leq k!\left(\frac{\as}{\lambda}\right)^{k+1}\bar b_{(k)} \\[10pt]
  \left(\frac{k!(\as/\lambda)^{k+1}\bar b_{(k)}}{|\Delta_k|}\right)^{n_c+1} &
  \mbox{ if }	|\Delta_k|>k!\left(\frac{\as}{\lambda}\right)^{k+1}\bar b_{(k)}
\end{array}
\right. \, 
\ee
and the credibility interval
\begin{align}
  \label{eq:intervalCHbar}
  d_k^{(p)}& =\left\{
  \begin{array}{l l} 
    k! \left(\frac{\alpha_s}{\lambda}\right)^{k+1} \bar{b}_{(k)} \frac{n_c+1}{n_c} p\% & 
    \text{if} \qquad p\% \leq \frac{n_c}{n_c+1} \\\\
    k! \left(\frac{\as}{\lambda}\right)^{k+1} \bar{b}_{(k)} \left[(n_c+1)(1-p\%)\right]^{(-1/n_c)} & 
    \text{if} \qquad p\% > \frac{n_c}{n_c+1}   \\
  \end{array}\right.\qquad .
\end{align}

The introduction of the $(n-1)!$ term in the expansion, which represents the main modification with respect to 
the original $\mathrm{CH}$ model, can be justified on the ground that such a factor is expected to appear 
in higher order perturbative calculations, e.g. those in the large-$\beta_0$ limit and in connection with 
renormalon contributions~\cite{PhysRevLett.73.1207,Zakharov:1992bx,Fischer:1997bs,Beneke:1998ui}. 

The optimal value for the rescaling factor $\lambda$ can be determined empirically by observing how the model 
fares in predicting MHOUs for observables for which higher order perturbative computations are available. 
In Section~\ref{sec:global} we will present such a determination of $\lambda$ from a study based on a comprehensive
set including more than thirty observables. This method of determining $\lambda$ brings some frequentist 
contamination into the Bayesian approach. We consider this drawback acceptable at the present stage, but 
we note that one could in principle further improve the model by introducing an additional prior for the value 
of $\lambda$ and thus avoid the frequentist contamination. The frequentist study on $\lambda$ performed 
in this work can then perhaps be used as a guide for the formulation of such an additional prior.

\subsubsection{Extension to hadronic observables}
\label{sec:had-observables}

The original CH model was formulated focusing on observables in processes without hadrons in the initial
state, and its extension to observables with initial state hadrons is potentially not straightforward. A generic hadronic 
observable (e.g. a total cross section) can be written as a convolution integral
\be
O_k (\tau,Q) = \mathcal{L}(Q) \otimes \sum_{n=l}^k \alpha_s^n C_n(Q)
\label{eq:genhadr}
\ee
where $\mathcal{L}$ is the parton-parton luminosity, $C_n(Q)$ is the hard-scattering 
coefficient function,
$\tau$ is an appropriate hadronic scaling variable, $Q$ is the characteristic energy scale of the process and 
$\otimes$ denotes a generic convolution in the space of the hadronic scaling variables (not explicitly shown 
on the right hand side of the equation). The unphysical renormalisation and factorisation scales are taken to 
be equal to $Q$ as in the non-hadronic case, and they are not explicitly shown.
In eq.~(\ref{eq:genhadr}), the perturbative coefficient functions $C_n$ are usually distributions, and not simple 
numbers like the coefficients $c_n$ in the perturbative expansion of the non-hadronic observables. 
This means that it is not possible to directly apply the \chbar\ method described in Section \ref{sec:chbarfac} to 
hadronic observables. This problem can be overcome in two ways.
\begin{enumerate}
\item A first approach is to express the hadronic observable as a series expansion whose coefficients 
include the convolution with the parton-parton luminosities, i.e. to rewrite eq.~(\ref{eq:genhadr}), in analogy with
the non-hadronic case, in the form
\be
O_k (\tau,Q) = \mathcal{L}(Q) \otimes \sum_{n=l}^k \alpha_s^n C_n(Q)
\equiv \sum\limits_{n=l}^{k}\left(\frac{\alpha_s}{\lambda_h}\right)^n (n-1)!~H_n(\tau,Q)
\label{eq:genhadr2}
\ee
where we have defined
\be
H_n(\tau,Q) \equiv  \frac{\lambda_h^n}{(n-1)!} h_n \equiv  \frac{\lambda_h^n}{(n-1)!} \mathcal{L}(Q) \otimes C_n(Q)\, .
\ee  
We now denote the rescaling parameter with $\lambda_h$, rather than $\lambda$, to stress the fact that its value is 
a priori potentially different from the one used in the case of non-hadronic observables.
We then proceed like in the non-hadronic case, submitting the expansion coefficients $H_n$ to the same Bayesian 
priors used in the non-hadronic case. 

This approach is based on the assumption that the contribution coming from the non-perturbative physics encoded 
in the parton-parton luminosity is roughly the same at each perturbative order, or more generally that its presence 
does not spoil the assumptions of the model. This approach has been adopted in some of the papers that have 
used the CH approach in its original formulation, e.g. \cite{Goria:2011wa, Forte:2013mda}.

\item A second approach is based on rewriting the observable in Mellin space, in the form 
  \be
  O_k(N,Q) = 
  \mathcal{L}(N+1) \sum_{n=l}^k \left(\frac{\alpha_s}{\lambda_h}\right)^n (n-1)!~B_n(N,Q)\, ,
  \ee
  where
  \be
  B_n(N,Q) \equiv \frac{\lambda_h^n}{(n-1)!} \int_0^1 dx\, x^{N-1}\, C_n(x,Q)
  \ee
is the Mellin transform of the short-distance coefficient function $C_n$, rescaled by the factor $\lambda_h^n/(n-1)!$ introduced 
in \chbar, and $\mathcal{L}(N+1)$ is the Mellin transform of the parton-parton flux.
We then observe that, if the Mellin inversion integral can be shown to be dominated by a single Mellin moment $O_k(N_0,Q)$, one can simply apply the Bayesian priors of the \chbar\ approach to the short-distance coefficients $B_n(N_0,Q)$,
which are ordinary numbers, and determine the uncertainty for the dominant moment series. This uncertainty can 
then be translated back to the uncertainty on the full result by an appropriate rescaling.
This approach is viable because one can show that at least in some cases (see e.g. \cite{Bonvini:2010tp,Bonvini:2012an})  
such a dominant Mellin moment exists and gives a good approximation to the full result. 

The main limitation of this approach, which a priori would be preferred because it eliminates the possible contamination 
due to non-perturbative physics, is that it relies on the predominance of not only a single Mellin moment but also a single 
production channel (e.g. gluon-gluon fusion in Higgs production at the LHC) at all orders. If this is not the case, the need 
to reweigh the various dominant Mellin moments in the different parton channels will reintroduce contamination from 
non-perturbative physics.
A second, practical, limitation is that perturbative results are rarely available in Mellin moment space from public codes, 
limiting the straightforward application of this method to very few cases.
Because of these limitations we use the first approach, i.e. the convolution, as our main tool in this paper, but we also present in Appendix~\ref{section:mellinAppendix} two case studies for the Mellin-moment method.
\end{enumerate}

\section{Global survey}\label{sec:global}

In this Section we assess the performance of the scale-variation procedure and of 
the \chbar\ approach by studying how well they estimate the MHOUs when applied to
a wide set of observables. For every observable in the set we consider two quantities:
\begin{enumerate}
\item the size of the uncertainty predicted  at a given perturbative order $k$ by the approach 
  under consideration;
\item the known perturbative result for the same observable at order $k+1$.
\end{enumerate}
For each of the methods we then determine its {\sl global success rate} in predicting the missing
higher order uncertainties at order $k$, defined as {\sl fraction} of observables for which the 
result of the calculation at order $k+1$ falls within the uncertainty interval predicted by the model 
for the order $k$ computation.

In the case of the scale-variation method we study the behaviour of the global success rate as we 
vary the scaling factor $r$ defined in Section~\ref{sec:th-sv}.
The observed success rate can then be used to assign an \emph{a posteriori} heuristic confidence level (CL) to the uncertainty 
intervals obtained with a given value of $r$.

In the case of the \chbar\ Bayesian approach we repeat the analysis described above for various values of 
$\lambda$ and, since we now have a probabilistic interpretation of the resulting uncertainty intervals, various Degrees of Belief (DoB). This allows us to determine the optimal value of $\lambda$ 
to be used in \chbar, defined as the value of $\lambda$ for which the 
model has a global success rate which is closest to the requested DoB, for every possible DoB.

\subsection{Setup}

We perform two separate analyses, one for observables in processes without hadrons in the initial state ({\em non-hadronic} 
observables) and one for observables in processes with hadrons in the initial state ({\em hadronic} observables).

The non-hadronic observables considered in our analysis are listed in Table~\ref{tab:non-hadronic-obs-text}. For each observable we show 
 the leading order in $\alpha_s$, the maximum known order in $\alpha_s$ and a reference to the original 
literature from which we have extracted the values of the perturbative coefficients.
When the leading order contribution for these observables is entirely electroweak in nature, we do not include the 
first coefficient $c_0$ in the analysis when using the \chbar\ approach, as was done in~\cite{Cacciari:2011ze}.  This is because we are interested in a 
perturbative expansion  in terms of the strong coupling.

\begin{table}[p]
  \small
  \centering
  \ra{1.1}
  \begin{tabular}{@{}cccc@{}}\toprule
    \multicolumn{4}{c}{\text{\LARGE Non-Hadronic observables }} \\
    Observable & Leading order in $\alpha_s$ & Highest known order in $\alpha_s$ & Reference \\
    \midrule
    $R = \frac{\sigma(e^+e^-\to\text{hadr})}{\sigma(e^+e^-\to\mu^+\mu^-)}$ & 0 & 3 & \cite{Baikov:2008jh}    \\
    \midrule
    Bjorken sum rule                & 0 & 3 & \cite{Larin:1990zw} \\
    GLS sum rule                     & 0 & 3 & \cite{Larin:1991tj} \\
    \midrule
    $\Gamma(b\to c e\bar{\nu}_{e})$   & 0 & 2 & \cite{Biswas:2009rb}       \\
    \midrule
    $\Gamma(Z\to\text{hadr})$       & 0 & 4 & \cite{Baikov:2012er}       \\
    $\Gamma(Z\to b\bar{b})$          & 0 & 3 & \cite{Chetyrkin:1994js}    \\
    \midrule
    3-jets Thrust                        & 1 & 3 & \cite{Weinzierl:2009yz}    \\
    3-jets Heavy jet mass          & 1 & 3 &                            \\
    3-jets Wide jet broadening  & 1 & 3 &                            \\
    3-jets Total jet broadening   & 1 & 3 &                            \\
    3-jets C parameter              & 1 & 3 &                            \\
    3-to-2 jet transition              & 1 & 3 &                            \\
    \midrule
    $\gamma_{ns}^{(+)}(N=2)$        & 1 & 3 & \cite{Larin:1996wd} \\
    $\gamma_{qq}(N=2)$                & 1 & 3 & \\
    $\gamma_{qg}(N=2)$                & 1 & 3 & \\
    \midrule
    $H\to b\bar{b}|_{m_b=0}$      & 0 & 4 & \cite{Baikov:2005rw}       \\
    $H\to gg$                               & 2 & 5 & \cite{Baikov:2006ch}       \\
    $H\to \gamma\gamma$         & 0 & 2 & \cite{Maierhofer:2012vv}   \\
    \bottomrule
  \end{tabular}
  \caption{List of non-hadronic observables used in the global survey. Note that when the leading term is purely electroweak the first coefficient, $c_0$, is not used when studying these non-hadronic observables in the Bayesian approach.}
  \label{tab:non-hadronic-obs-text}
\end{table}

\begin{table}[p]
  \small
  \centering
  \ra{1.1}
  \begin{tabular}{@{}cccl@{}}\toprule
    \multicolumn{4}{c}{\text{\LARGE Hadronic observables }} \\
    Observable & Leading order in $\alpha_s$  & Highest known order in $\alpha_s$ & Reference  \\
    \midrule
    $pp\to H$                                 & 2 & 4 & HIGLU \cite{Spira:1995rr,Spira:1995mt} \\
    $pp \to b\bar{b}\to H$                        & 0 & 2 & bbh@nnlo \cite{Harlander:2003ai} \\
    $pp\to t\bar{t}$                         & 2 & 4 & top++ \cite{Czakon:2013goa} \\
    \midrule
    $pp\to Z \to e^+e^-$                              & 0 & 2 & DYNNLO \cite{Catani:2009sm} \\
    $pp\to W^+ \to e^+\overline{\nu}_e$     & 0 & 2 & DYNNLO  \\
    $pp\to W^- \to e^-\nu_e$                       & 0 & 2 & DYNNLO  \\
    $pp\to Z^* \to ZH $                                & 0 & 2 & vh@nnlo \cite{Brein:2003wg} \\
    $pp\to W^{\pm *} \to W^{\pm}H $          & 0 & 2 & vh@nnlo \\
    \midrule
    $pp\to b\bar{b}$                         & 2 & 3 & MCFM \cite{Campbell:1999ah,Campbell:2002tg} \\
    $pp\to Z+\mathrm{j}$                 & 1 & 2 & MCFM \\
    $pp\to Z+2\mathrm{j}$               & 2 & 3 & MCFM \\
    $pp\to W^\pm+\mathrm{j}$        & 1 & 2 & MCFM \\
    $pp\to W^\pm+2\mathrm{j}$      & 2 & 3 & MCFM \\
    $pp\to ZZ$                                 & 0 & 1 & MCFM \\
    $pp\to WW$                               & 0 & 1 & MCFM \\
    \bottomrule
  \end{tabular}
  \caption{List of hadronic observables used in the global survey.}
  \label{tab:hadronic-obs-text}
\end{table}

The observables included in our hadronic analysis are listed in Table~\ref{tab:hadronic-obs-text}, where again we show the
leading order in $\alpha_s$, the highest known order in $\alpha_s$ and a reference to the code implementing the computation
that we used to evaluate the perturbative coefficients. 
In this case, the leading order coefficient (i.e. the first one) is always retained for the analysis with the \chbar\ approach, 
independently of its perturbative order in the strong coupling\footnote{When this first coefficient is of zeroth order we 
set the $(n-1)!$ term equal to one in the perturbative expansion in  eq.~(\ref{eq:genhadr2}).}. In order to avoid biasing the analysis by using different parton 
distribution functions (PDFs) at different orders, we always use the same NNLO PDFs for all perturbative orders, with 
the exception of the scale-variation study shown in the right plot of Figure~\ref{fig:globalhadronicSV}.

All the coefficients and the specific parameters for the calculations are given in Appendix~\ref{sec:tables}, 
in Tables \ref{tab:non-hadronic-obs} and \ref{tab:hadronic-obs}. For all our analyses, we have used a private 
Mathematica code.

\subsection{Results}

\subsubsection{Scale Variation}

In this Section, we study the performance of the standard scale-variation approach. An outcome 
of this analysis is the determination of a heuristic confidence level (CL) for the uncertainty intervals given by scale 
variation, as a function of the scaling factor $r$ that sets the range over which the scales are varied, 
$\mu \in [Q/r, rQ]$.

\begin{figure}
  \centering
  \includegraphics[width=0.48\textwidth]{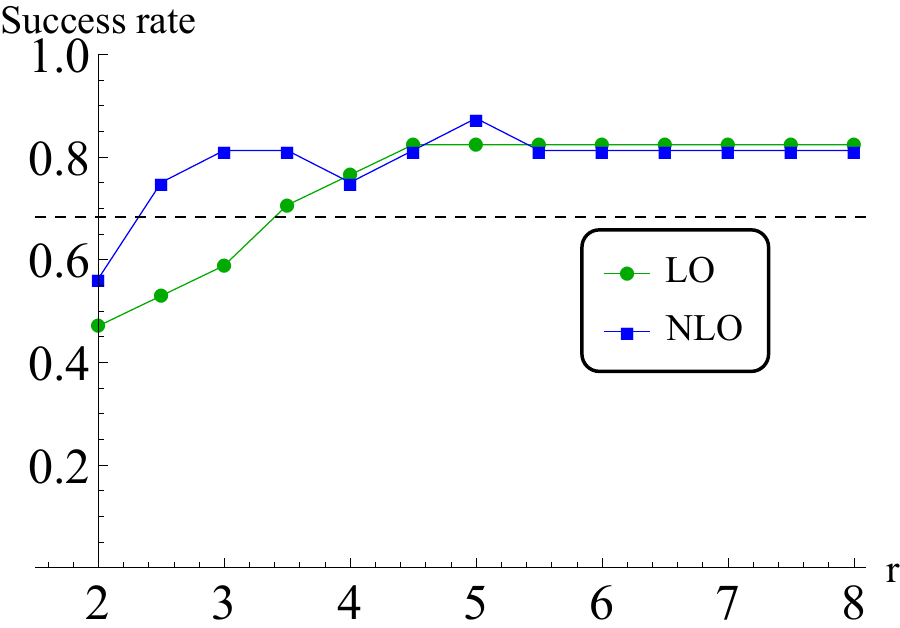}
  \includegraphics[width=0.48\textwidth]{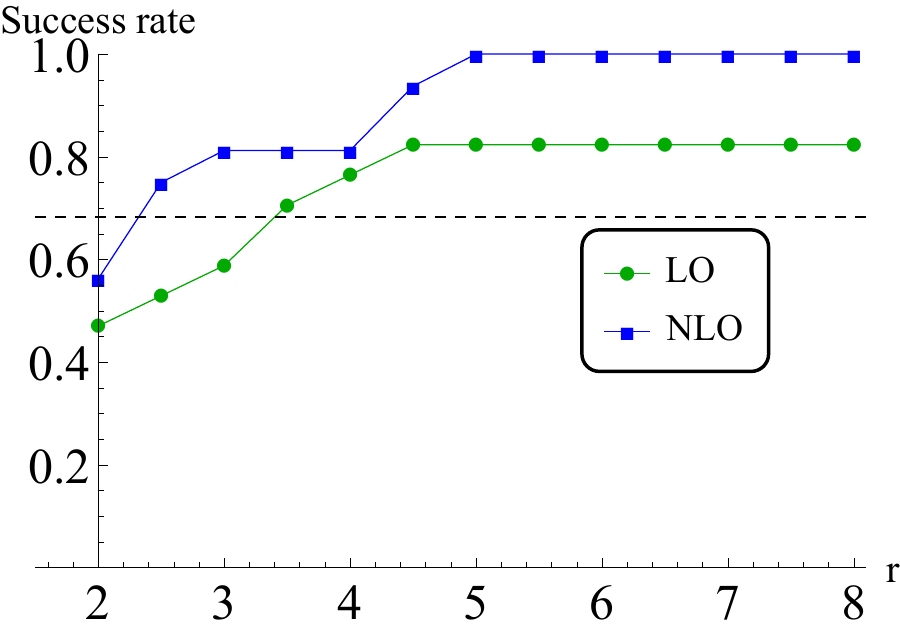}
  \caption{Fraction of observables whose known higher order is found to be contained within the uncertainty interval given by  scale variation
    between $\mu=Q/r$ and $\mu=r Q$. Left plot: only the extremes and the central value of the $[Q/r,r Q]$ are used. Right plot: the full $[Q/r,r Q]$ interval is scanned.
}
  \label{fig:SV}
\end{figure}

In the non-hadronic case, we also compare two of the prescriptions given in Section~\ref{sec:th-sv}, 
which are supposedly  the most widely used ones: 
a) take the maximum and the minimum of the cross sections obtained with $\mu=rQ$ or $\mu = Q/r$, 
as explained in eq.~(\ref{eq:scale1});
b) take the maximum and the minimum while scanning the whole interval of scales between $Q/r$ and 
$rQ$, as explained in eq.~(\ref{eq:scale2}).
Results for the first prescription (i.e. using only the extreme values) are given in the left plot of Figure~\ref{fig:SV}.
At LO, the heuristic CL of the scale-variation uncertainty intervals for the conventional $r=2$ 
value is of the order of 50\%, and it reaches a 68\% level for $r$ close to $4$. For larger values of $r$, the CL 
stabilises around 80\%.
At NLO, the CL is still of the order of 50\% at $r=2$, but it increases more rapidly with $r$ than at LO, and it 
is already around 68\% for $r\simeq 2.5-3$. For higher values of $r$ it stabilises around $80\%$.
Results for the second scale-variation prescription (i.e.~doing a full scan) are given in the right plot of Figure~\ref{fig:SV}.
While the LO results are identical to those of the first prescription, the NLO heuristic CLs are significantly larger for $r\geq 4$, reaching $100\%$ at $r=5$.
This is likely explained by the fact that, being the scale variation  of an observable calculated to NLO accuracy 
usually non monotonic, a full scan can capture better its overall variation than the evaluation of  two or three fixed points only.

\begin{figure}
  \includegraphics[width=0.48\textwidth]{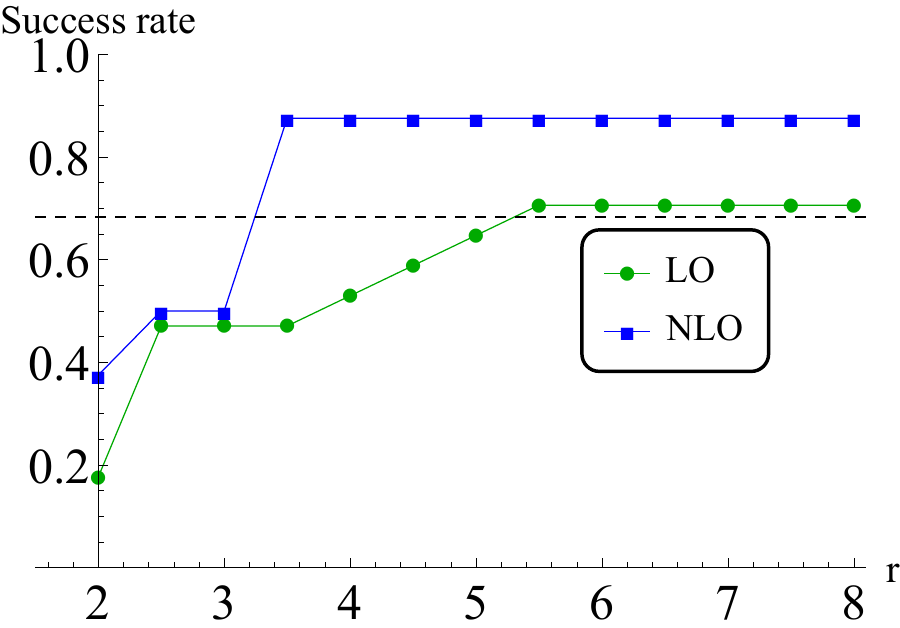}\quad
  \includegraphics[width=0.48\textwidth]{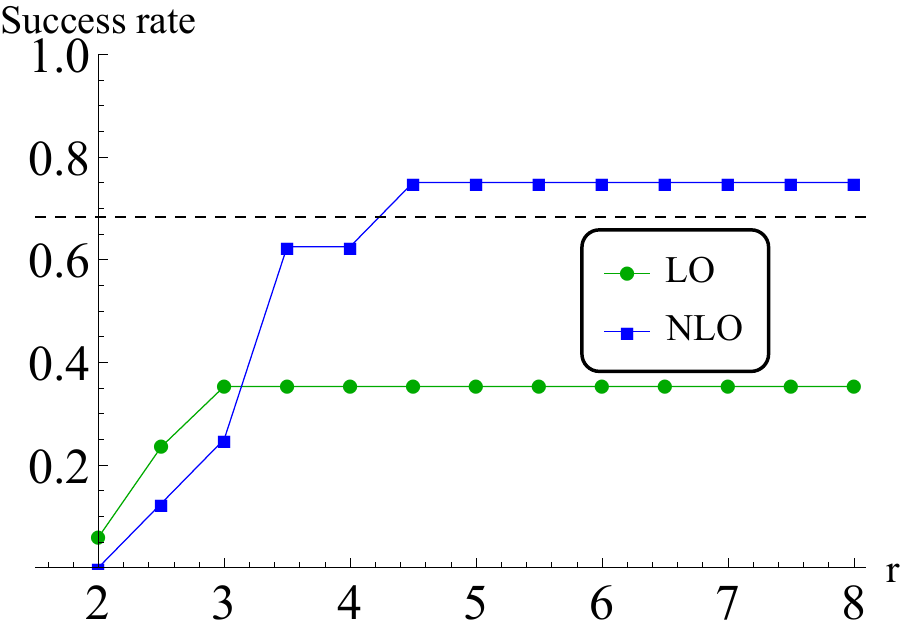}
  \caption{Fraction of observables whose known higher order is found to be contained within the uncertainty interval given by  renormalisation and factorisation scale variation
    between $\mu_{r,f}=Q/r$ and $\mu_{r,f}=r Q$ with the constraint $1/r \le \mu_r/\mu_f \le r$.
Only the seven points at the extremes and at the centre of the scale-variation interval are used.
Left plot: NNLO-evolved PDFs are used with all perturbative orders. Right plot: PDFs evolution order is matched with the perturbative order of the observable.
}
  \label{fig:globalhadronicSV}
\end{figure}

We have also examined scale-variation uncertainties in the case of hadronic observables. Since hadronic 
cross sections depend on two scales, the factorisation and renormalisation scale, we vary them independently 
to obtain the scale-variation interval. As often done in literature, we do not perform a full scan (too computationally 
expensive) but rather evaluate the observables only at the centre and at the extremes of a scale range, avoiding 
combinations that generate large logarithms, as explained at the end of Section~\ref{sec:th-sv}.
Figure~\ref{fig:globalhadronicSV} shows the results of the analysis of the full set of hadronic observables. 
We have calculated the cross sections both using NNLO PDFs at each order (left plot) and using order-matched 
PDFs (right plot), i.e. using LO PDFs for the LO computation, NLO ones at NLO, etc\footnote{We have used NNPDF 2.1~\cite{Ball:2011uy} at LO, and NNPDF 2.3~\cite{Ball:2012cx} at NLO and NNLO.}.  At each perturbative order, 
the two choices are equivalent up to higher order terms. 
In both cases, we see that, as common wisdom dictates, the LO scale-variation uncertainty fails to capture the size of 
the NLO correction. At NLO the two prescriptions differ qualitatively in their performance. When using always NNLO PDFs we 
can associate a $40\%$ heuristic CL to the standard scale variation with $r=2$. The $68\%$ CL level is attained around 
$r=3$, and the CL then stabilises around $90\%$ CL for $r\geq 3.5$. When using order-matched PDFs, on the other hand, we 
obtain very small heuristic CL (less than $30\%$) for $r\leq 3$. The CL reaches $68\%$ for $r$ just over $4$ and then 
stabilises around $80\%$ for larger values of $r$.
These two analyses for hadronic observables suggest that in the scale-variation approach one may wish to use a rescaling factor $r\sim 3-4$ in order to obtain 
a reasonably conservative uncertainty interval, with a heuristic CL at least as large as 68\%.

\subsubsection{The modified Cacciari-Houdeau model \texorpdfstring{($\overline{\mathrm{CH}}$)}{CH-BAR}}

For each of the sets of observables listed in Tables~\ref{tab:non-hadronic-obs-text} and 
\ref{tab:hadronic-obs-text} we have  performed an analysis of the performance of the 
\chbar\ model in estimating the MHOUs. In this case, a parameter of the model is the  $\lambda$ (or $\lambda_h$
factor) that defines the effective expansion parameter of the perturbative series as written in the model, 
see eq.~(\ref{eq:chbarexp}) and eq.~(\ref{eq:genhadr2}). As far as the size of the uncertainty intervals is concerned, the parameter $\lambda$ (or $\lambda_h$)
plays a role analogous to that of $r$ in the scale-variation approach: the final result will depend on its value. However, since in the Bayesian model the widths of the 
uncertainty intervals are associated with properly defined credibility values, one can explicitly 
determine the optimal value for $\lambda$ by requiring that the model performs as expected, i.e. that the observed global success rate corresponds to the DoB of the uncertainty intervals used in the analysis.

\begin{figure}[t]
  \centering
  \includegraphics[width=0.42\textwidth,trim=0 -1.1cm 0 0]{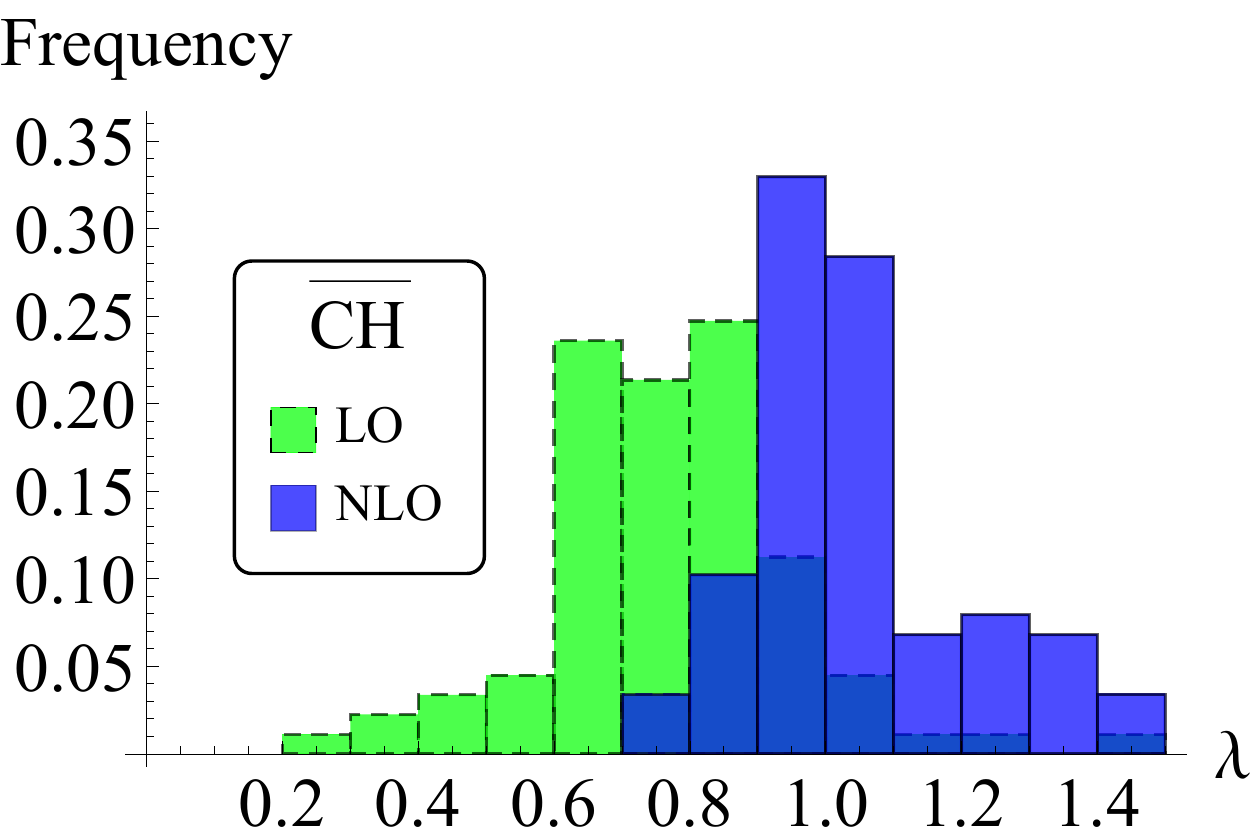}\quad
  \includegraphics[width=0.55\textwidth]{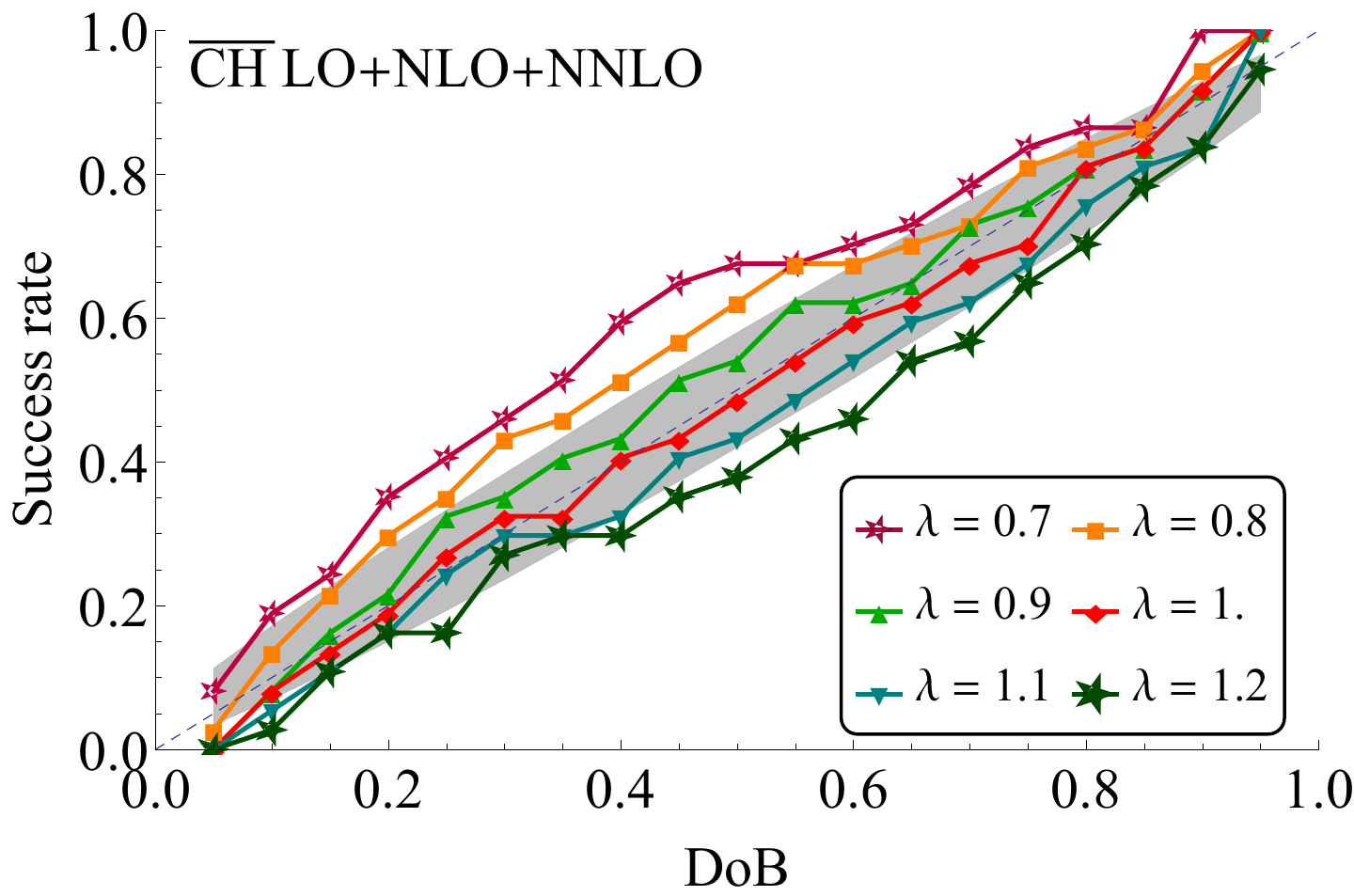}
  \caption{Non-hadronic survey: comparisons between DoB and actual success rate, to determine 
    the most appropriate value for $\lambda$. Left, histogram of the optimal $\lambda$ value 
    obtained with a DoB scan. Right, plot of the success rate vs the requested DoB for six values 
    of $\lambda$.}
  \label{fig:CHApproxFacAll}
\end{figure}

\begin{figure}[t]
  \includegraphics[width=0.42\textwidth,trim=0 -1.1cm 0 0]{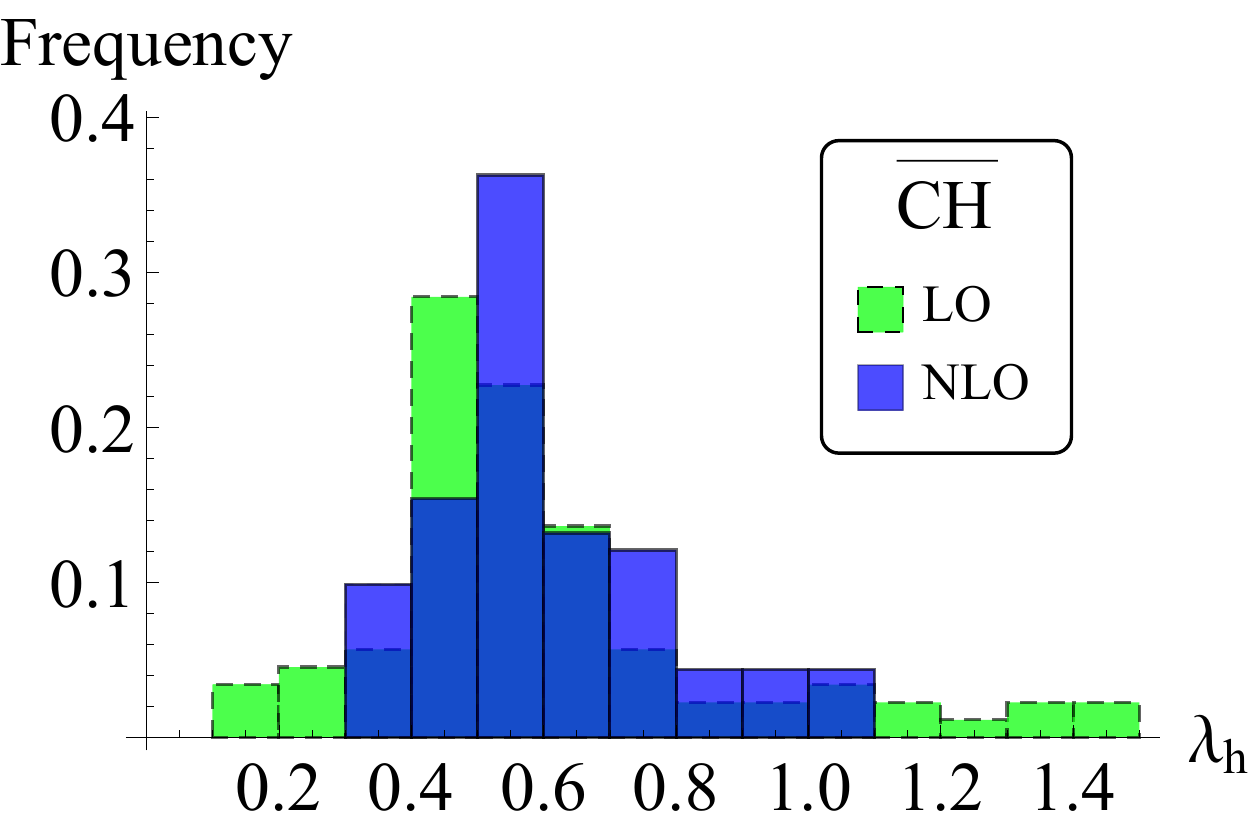}\quad
  \includegraphics[width=0.55\textwidth]{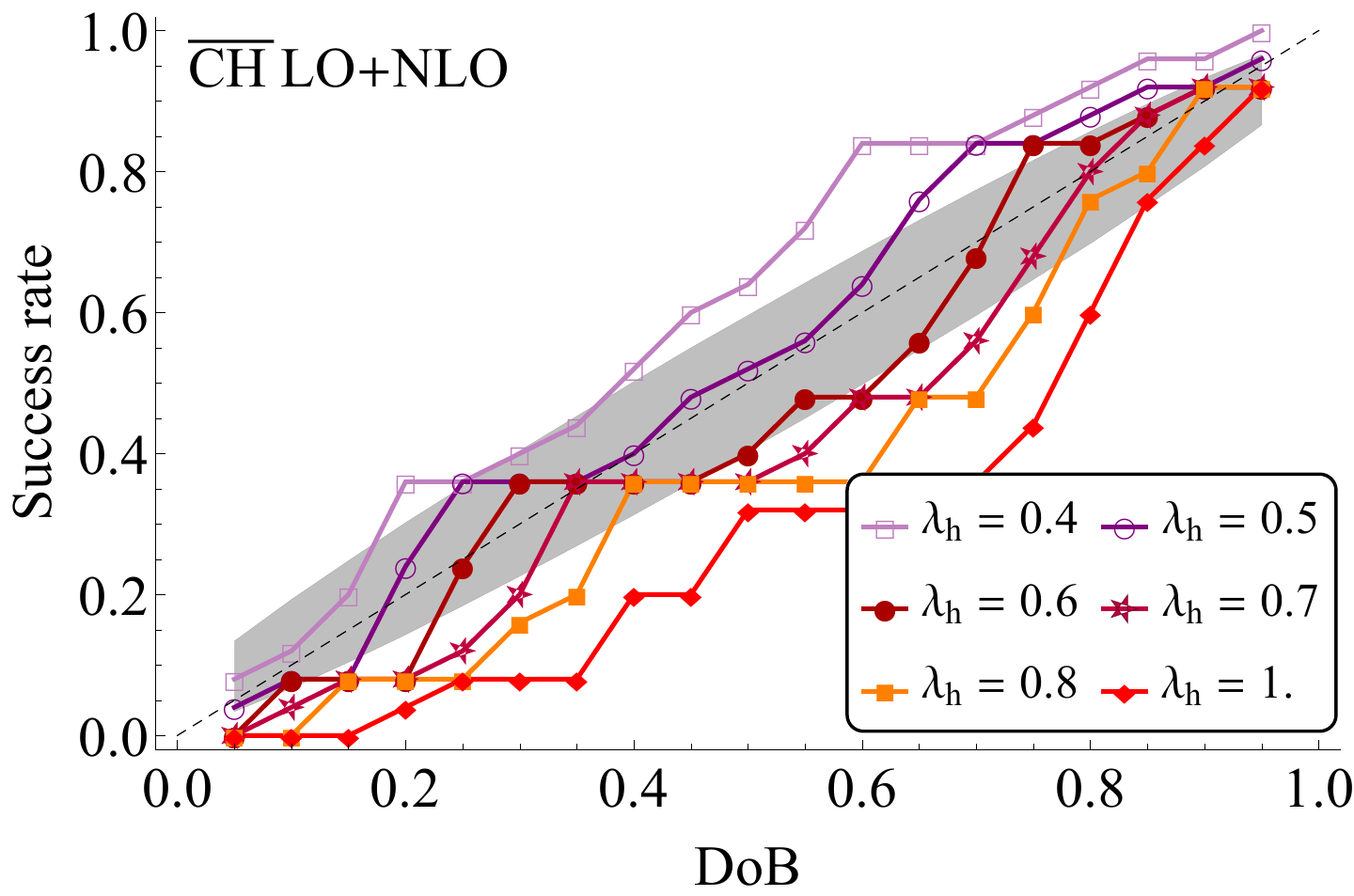}
  \caption{Hadronic survey: comparison between DoB and actual success rate to determine the most 
    appropriate value for $\lambda_h$ for all hadronic observables.}
  \label{fig:globalhadronicNoDIS}
\end{figure}

We first study the non-hadronic case. We show the results of this analysis graphically in Figure~\ref{fig:CHApproxFacAll} 
in two different and complementary ways. Both analyses use observables calculated at perturbative orders ranging from LO to N$^3$LO, for a total of 37 tests performed using the numerical coefficients given in Table~\ref{tab:non-hadronic-obs} in Appendix~\ref{sec:tables}.
The histogram in Figure~\ref{fig:CHApproxFacAll}~(left) is obtained by varying the DoB between 0.05 
and 0.95 in steps of 0.01 (the uncertainty interval returned by \chbar\ varies of course accordingly). For each DoB value, we determine the $\lambda$ value which gives 
the best agreement with the condition DoB = global success rate. The resulting $\lambda$ values are 
plotted in a histogram. 
At LO the preferred values for $\lambda$ can be seen to be between $0.6$ and $0.9$, while at NLO
the histogram shows a preference for the range $0.9$~-~$1.1$.
The plot in Figure~\ref{fig:CHApproxFacAll}~(right) shows instead how DoB and success rate 
compare for different values of $\lambda$ in a global analysis of LO, NLO and NNLO observables. 
We see that for values of $\lambda$ in the $0.9$~-~$1.1$ range the requested DoB agrees well with 
the observed success rate of the uncertainty prediction.\footnote{This frequentist-like determination of $\lambda$ is itself subject to an uncertainty due to 
the finite size of the set of observables that we have used, which results in a statistical 
error on the observed success rate (see Appendix~\ref{section:statuncapp} for a quantitative 
analysis). This statistical error is displayed as a grey band in Figure~\ref{fig:CHApproxFacAll}
(right). One can see how it roughly translates into a limiting precision of  $\pm 0.2$ in the 
determination of $\lambda$.
} This is in agreement with the result 
that we obtain from the histogram analysis.

We perform the same analysis for the hadronic observables set using the coefficients given in Table~\ref{tab:hadronic-obs} in Appendix~\ref{sec:tables}.
Figure~\ref{fig:globalhadronicNoDIS}~(left) shows the histogram of the optimal values of $\lambda_h$ 
for the DoB scan made using the full set of hadronic observables. The histogram peaks around 
$\lambda_h \simeq 0.5$ at NLO, which is smaller than the preferred $\lambda$ value obtained from the analysis of 
non-hadronic observables.
In Figure~\ref{fig:globalhadronicNoDIS}~(right) we plot the success rate as a function of the DoB 
of the $\overline{\mathrm{CH}}$ intervals for various values of $\lambda_h$ for all hadronic observables 
at LO and NLO. We observe that the preferred value of $\lambda_h$ oscillates
between $0.5$ and $0.6$ according to the requested DoB. Since we are mainly interested in determining
68\% and 95\% DoB intervals, we choose a value of $\lambda_h$ equal to $0.6$ as our best estimate,
since it appears to be the one for which the model performs better in this DoB range.

The results of the analyses presented in this section allow us to define the optimal values for the parameters
in the $\overline{\mathrm{CH}}$ model as follows: we use a parameter $\lambda = 1$ when considering 
non-hadronic observables,
while we  use $\lambda_h = 0.6$ when considering hadronic observables.\footnote{It may be tempting 
to speculate that the smaller value of $\lambda$ in the hadronic case (and therefore a larger effective expansion 
parameter for the series) may be explained by the generally larger number of gluons involved in these processes, 
and therefore by an expansion parameter closer to $\alpha_s C_A$ than to $\alpha_s C_F$, but we will refrain 
from doing so.}

\section{Benchmark processes}
\label{sec:benchmark}
In this Section, we compare the results obtained when computing MHOUs using either the \chbar\ or the 
scale-variation prescription for a set of benchmark processes that we consider interesting either 
because they provide an ideal testing ground for the \chbar\ method ($e^+e^-\to$~hadrons and the Higgs 
decay into two gluons) or are particularly relevant for LHC phenomenology (electroweak vector boson, top 
quark and Higgs production, Higgs decay into two photons).
 
We use the results obtained in the global survey (see Section~\ref{sec:global}) to fix the parameters of the 
models. We recall that, for the \chbar\ model, in the case of observables without initial-state hadrons the 
preferred value  is $\lambda$ = $1$, while for observables involving initial-state hadrons it is $\lambda_h = 0.6$.

For each process, we compare the uncertainty intervals obtained from the scale-variation procedure with $r=2$ 
and $r=4$ with the $68\%$ and  $95\%$ DoB intervals obtained using \chbar. When analysing the \chbar\ results, we 
also consider the behaviour of the posterior density function for the remainder of the series $\Delta_k$ when increasing 
the perturbative order. We show how, in most cases, the inclusion of further information leads to a progressive narrowing 
of the distribution and a consequent reduction of the uncertainty.

\subsection{Processes without hadrons in the initial state}

We first consider three processes without hadrons in the initial state: the total cross section for the production 
of hadrons in $e^+e^-$ collisions and the decay of a Standard Model Higgs boson into a pair of gluons or a pair 
of photons.

\begin{figure}[p]
	\centering
	\includegraphics[width=0.7\textwidth]{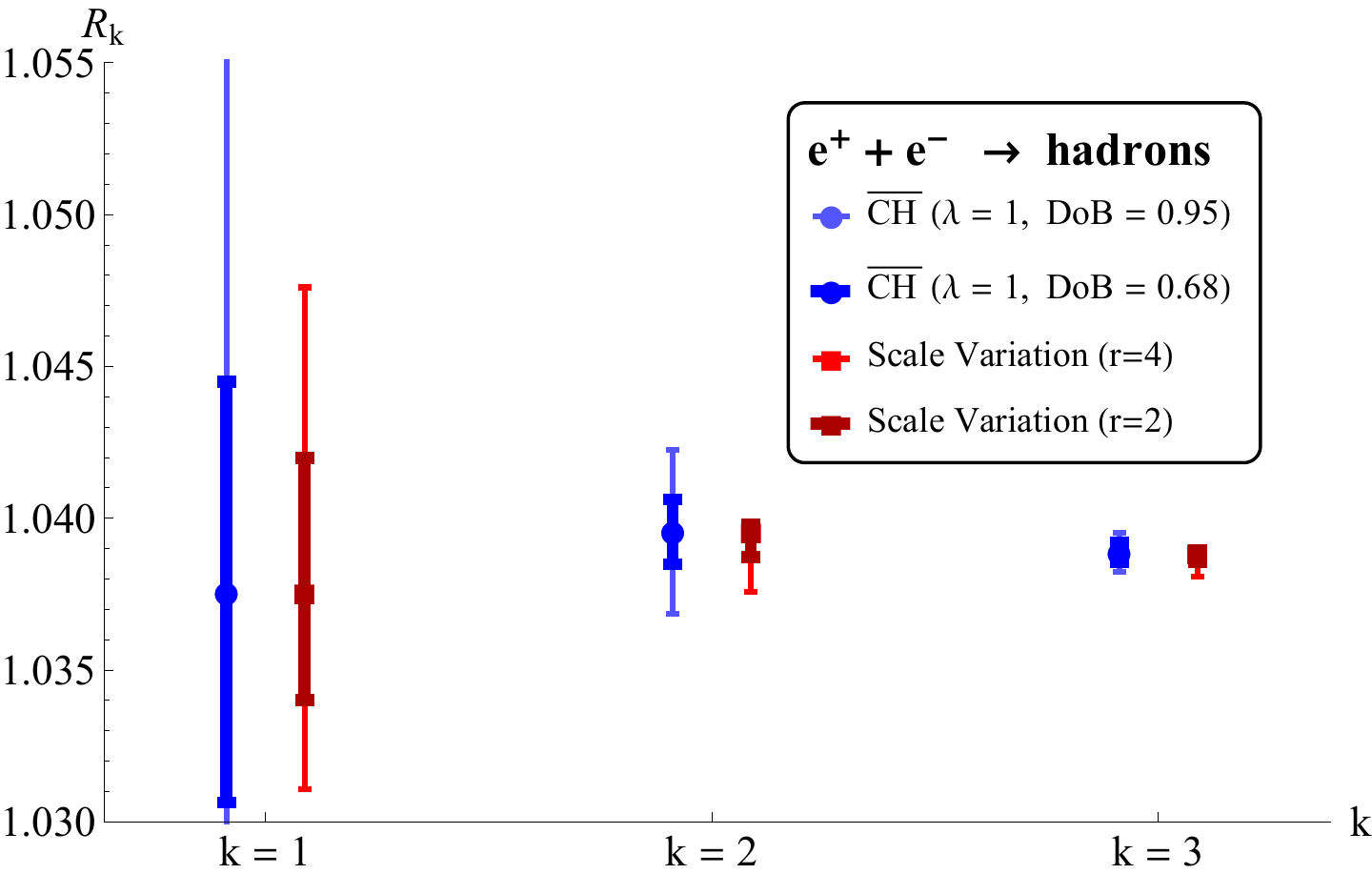}
	\caption{Size of the MHO uncertainty intervals at LO, NLO and NNLO for the $e^+e^-\to$~hadrons process at the $Z$ pole with the \chbar\ model with $\lambda=1$, compared to those predicted by scale
	variation.}
	\label{fig:epembars}
\end{figure}

\begin{figure}
	\centering
	\includegraphics[width=0.9\textwidth]{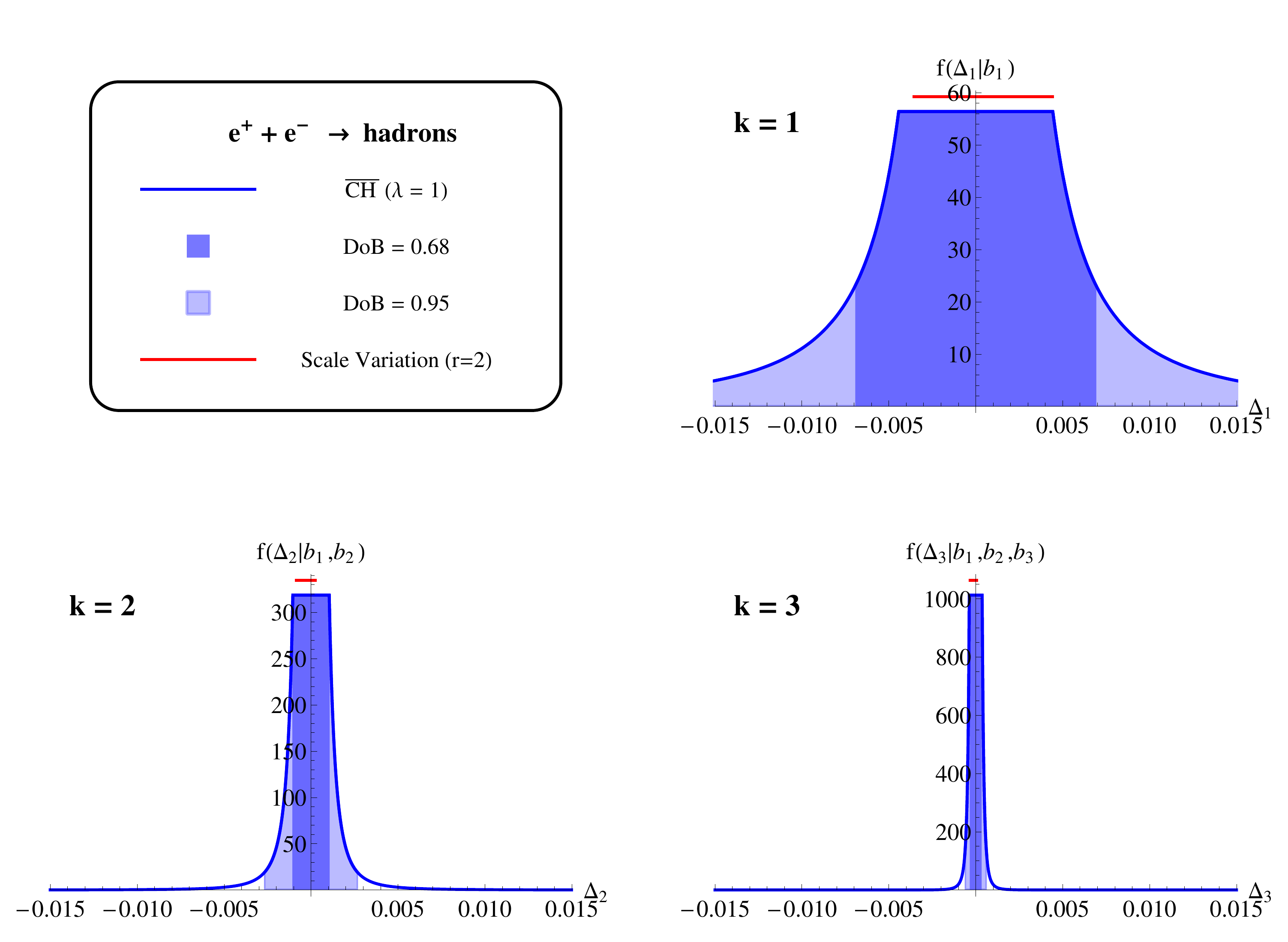}
	\caption{Posterior distribution for the remainder $\Delta_k$ (blue solid) for the $e^+e^-\to$~hadrons process at the $Z$ pole with the \chbar\ model,  $68\%$ DoB interval (blue fill), $95\%$ DoB interval (light-blue fill), scale-variation interval with $r=2$ (red solid)}
	\label{fig:epemposterior}
\end{figure}

As discussed in~\cite{Cacciari:2011ze}, the total cross section for $e^{+}e^{-} \to \text{hadrons}$ is an ideal testing 
case for understanding the behaviour of the \chbar\ model, as perturbative coefficients up to order $\alpha_s^3$ 
are available in the literature. Their numerical values are listed in Table~\ref{tab:non-hadronic-obs} in Appendix~\ref{sec:tables}.
In Table~\ref{subtable:hadroproduction}, we summarise the results of our study, comparing the size of the 68\% and
95\% DoB intervals obtained with the \chbar\ method with the uncertainty interval of the 
scale-variation procedure for $r=2$. A graphical representation of these intervals is shown in Figure~\ref{fig:epembars}.
These results show that 68\% DoB intervals from \chbar\ are always larger than scale-variation
intervals for $r=2$ and, especially at higher orders, agree better in size with those obtained using scale-variation with $r=4$.
In Figure~\ref{fig:epemposterior} we plot the full posterior distribution for the remainder of the perturbative
expansion, $\Delta_k \equiv\sum_{n=k+1}^\infty$, at each order $k$. We highlight the regions that contribute to the 68\% and 95\% DoB intervals and compare
them to the $r=2$ scale-variation intervals, showing how, in this case, the latter is always contained in the flat part
of the Bayesian credibility distribution for $\Delta_k$.

\begin{table}
\centering

\caption{\label{tab:resnonhadr}
Results for the analysis of missing higher order uncertainties for benchmark processes without hadrons in the initial state. We quote the perturbative order $k$ at which the observable is calculated, 
the central value for the theoretical prediction at that order, the MHOs uncertainty intervals computed using the \chbar\ model 
at 68\% DoB and 95\% DoB, and the uncertainty interval obtained using the scale variation (SV) procedure with $r=2$.}

\subtable[MHOUs for  hadron production in $e^+e^-$ collisions at the $Z$ pole. The perturbative series of the 
observable at the order $k$ is defined as 
$R_k( e^+e^- \rightarrow Z \rightarrow \mathrm{hadrons} ) = \sum_{n=0}^{k} \alpha_s^n c_n$. $R_0$ 
is normalised to 1.]
{
  \label{subtable:hadroproduction}
  \centering
  \ra{1.1}
  \begin{tabular}{@{}lcccc@{}}\toprule
  \multicolumn{5}{c}{\text{\LARGE $e^+e^- \rightarrow Z \rightarrow \mathrm{hadrons}$ }} \\
  \toprule
  Order &  $R_k$ & \chbar$_\mathrm{68\% DoB}$ & \chbar$_\mathrm{95\% DoB}$ & $\mathrm{SV}_{r=2}$ \\ \midrule
  \vspace{.2cm}
  $k = 1$ &  1.03756	& $\pm 0.00693$ & $\pm 0.04432$  & {\Large $\substack{+0.0044\\-0.0035}$} \\
  \vspace{.2cm}
  $k = 2$ & 1.03955	& $\pm 0.00107$ & $\pm 0.00270$  & {\Large $\substack{+0.00025\\-0.00084}$} \\
  \vspace{.2cm}
  $k = 3$ & 1.03887	& $\pm 0.00034$ & $\pm 0.00063$  & {\Large $\substack{+0.00006\\-0.00032}$} \\
  \bottomrule
  \end{tabular}
}

\subtable[MHOUs for Higgs decay into two gluons. The perturbative series of the observable at the order $k$ 
is defined as $\Gamma_k(H \to gg) = \sum_{n=2}^{k} \alpha_s^n c_n$.]
{
  \label{subtable:Hgg}
  \centering
  \ra{1.1}
  \begin{tabular}{@{}lcccc@{}}\toprule
  \multicolumn{5}{c}{\text{\LARGE $H \rightarrow gg$ }} \\
  \toprule
  Order &  $\Gamma_k$[MeV] & \chbar$_\mathrm{68\% DoB}$ & \chbar$_\mathrm{95\% DoB}$ & $\mathrm{SV}_{r=2}$	 \\ \midrule
  \vspace{.2cm}
  $k = 2$ & 0.185 & $\pm 0.065$	& $\pm 0.420$ & {\Large $\substack{+0.044\\-0.032}$} \\
  \vspace{.2cm}
  $k = 3$ & 0.305 & $\pm 0.041$ & $\pm 0.105$ & {\Large $\substack{+0.040\\-0.035}$} \\
  \vspace{.2cm}
  $k = 4$ & 0.342 & $\pm 0.017$ & $\pm 0.031$ & {\Large $\substack{+0.012\\-0.019}$} \\
  \vspace{.2cm}
  $k = 5$ & 0.345 & $\pm 0.009$ & $\pm 0.015$ & {\Large $\substack{+0.0004\\-0.006}$} \\
  \bottomrule
  \end{tabular}
}

\subtable[MHOUs for Higgs decay into two photons. The perturbative series of the observable at the order $k$ 
is defined as $\Gamma_k(H \to \gamma\gamma) = \sum_{n=0}^{k} \alpha_s^n c_n$.]
{
  \label{subtable:Hgammagamma}
  \centering
  \ra{1.1}
  \begin{tabular}{@{}lcccc@{}}\toprule
  \multicolumn{5}{c}{\text{\LARGE $H \rightarrow \gamma\gamma$ }} \\
  \toprule
  Order &  $\Gamma_k$[KeV] & \chbar$_\mathrm{68\% DoB}$ & \chbar$_\mathrm{95\% DoB}$ & $\mathrm{SV}_{r=2}$ \\ \midrule
  \vspace{.2cm}
  $k = 1$ & 9.548 & $\pm 0.030$	& $\pm 0.192$ & {\Large $\substack{+0.019\\-0.015}$} \\
  \vspace{.2cm}
  $k = 2$ & 9.556 & $\pm 0.004$ & $\pm 0.011$ & {\Large $\substack{+0.001\\-0.003}$} \\
  \bottomrule
  \end{tabular}
}

\end{table}

At the LHC, Higgs decay rates constitute one of the most important processes which do not involve initial-state hadrons.
Their precise knowledge is crucial for the extraction of the Higgs couplings to the other SM particles.
For our study we consider two Higgs decay channels which present complementary characteristics with respect to our theoretical knowledge
 and to their phenomenological relevance.
The first one is the Higgs decay into two gluons which, despite being not relevant for Higgs phenomenology at the LHC because 
of the large irreducible background due to QCD jets, is especially well suited as a test case for our Bayesian analysis. 
Indeed its perturbative QCD expansion is known up to N3LO, and QCD corrections are quite sizeable.
Next we study the decay of a Higgs boson into two photons. In this case, QCD corrections are 
rather small and its perturbative expansion is only known up to NLO. On the other hand, due to its clean experimental 
signature, it is of great phenomenological importance and it is indeed one of the channels where signs of new physics 
beyond the Standard Model are expected to appear. Again, numerical values for the coefficients of the perturbative 
expansions of these observables are collected in Table~\ref{tab:non-hadronic-obs} in Appendix~\ref{sec:tables}, while the results of our analysis are 
summarised in Tables~\ref{subtable:Hgg}-\ref{subtable:Hgammagamma}.

\begin{figure}[p]
	\centering
	\includegraphics[width=0.7\textwidth]{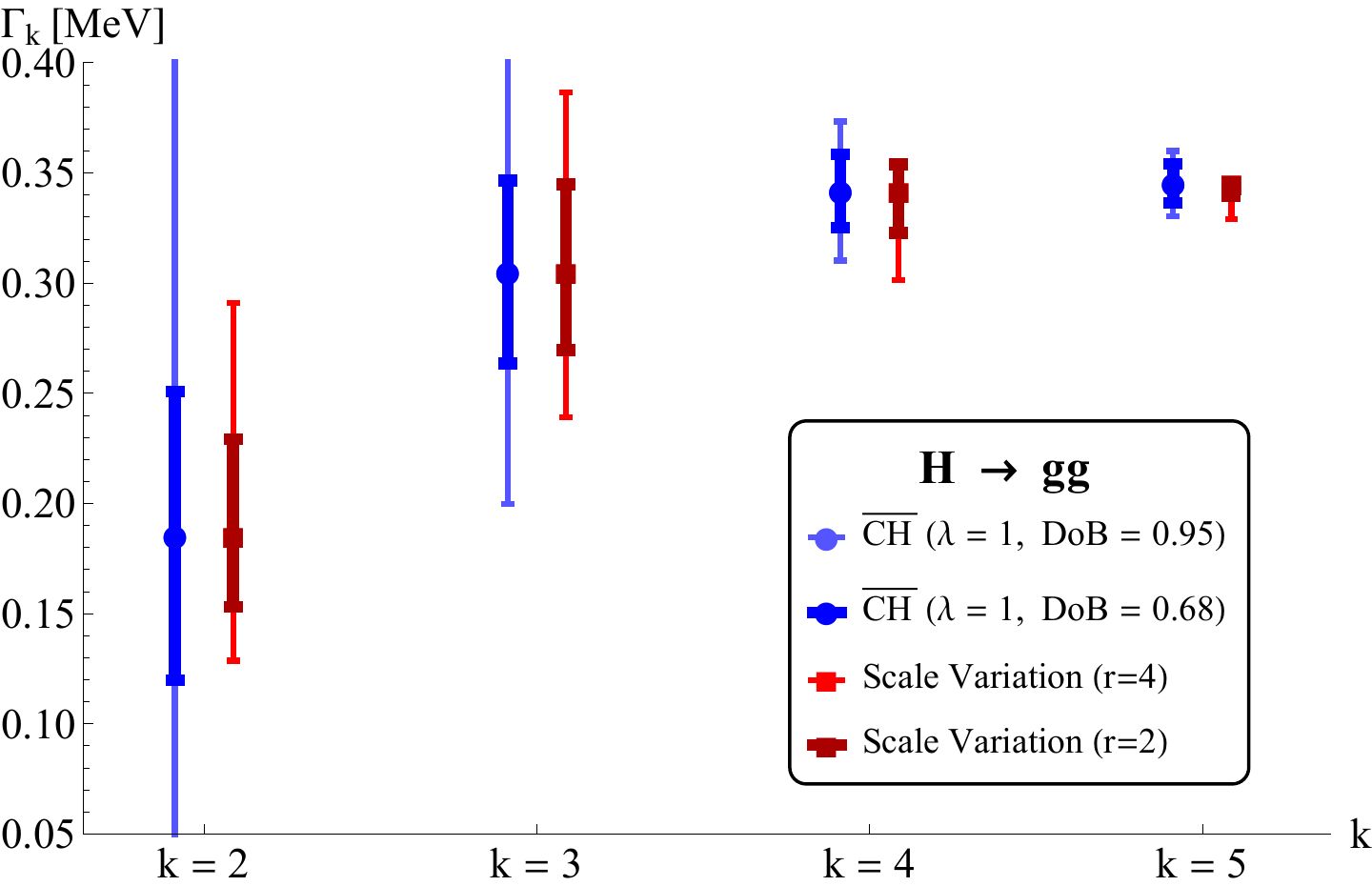}
	\caption{Size of the MHO uncertainty intervals at LO, NLO, NNLO and N3LO for the $H\to gg$ process  with the \chbar\ model with $\lambda=1$, compared to those predicted by scale
	variation.}
	\label{fig:hggbars}
\end{figure}

\begin{figure}
	\centering
	\includegraphics[width=0.8\textwidth]{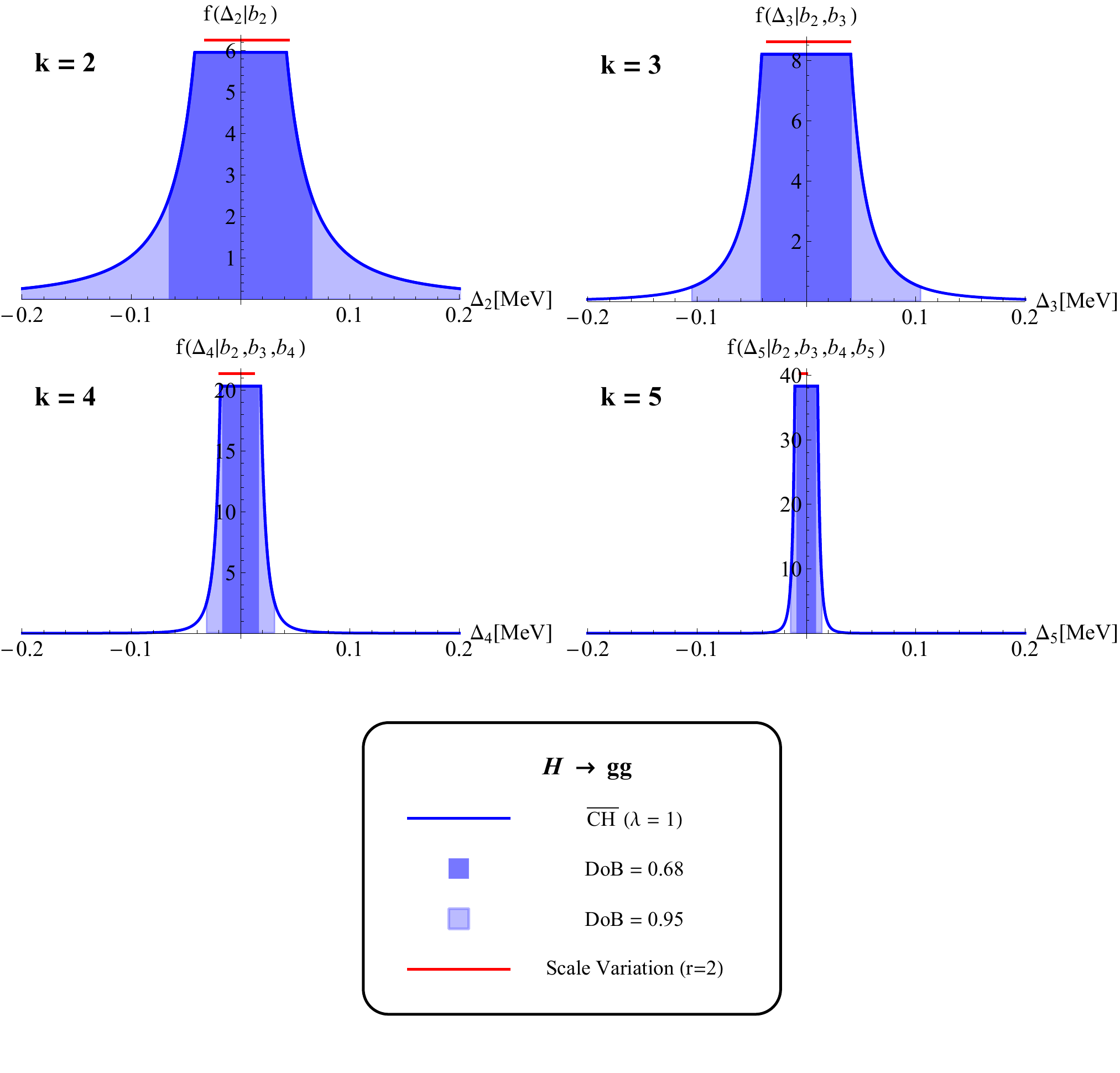}
	\caption{Posterior distribution for the remainder $\Delta_k$ (blue solid) for the $H\to gg$ process with the \chbar\ model, $68\%$ DoB interval (blue fill), $95\%$ DoB interval (light-blue fill), scale-variation interval  with $r=2$ (red solid).}
	\label{fig:hggposterior}
\end{figure}

For the $H \to gg$ process, we plot the uncertainty intervals obtained using the scale-variation and the \chbar\ methods in 
Figure~\ref{fig:hggbars}. 
We observe that the size of the $68\%$ DoB intervals obtained with the \chbar\ model is, with the exception of the N3LO band, slightly
bigger than the $r=2$ and smaller than the $r=4$ scale-variation intervals, coherently with what we have observed in the global survey.
We note here that, possibly because of a large NLO $K$-factor, neither the $r=2$ and $r=4$ scale-variation interval nor the 
$68\%$ DoB \chbar\ interval at LO contains the NLO result. Conversely at higher orders, where successive perturbative
corrections decrease in size, the next order result is always included in both the $68\%$ DoB and the $r=2$ intervals.
The posterior density distributions for $\Delta_k$ are plotted in Figure~\ref{fig:hggposterior}. We notice that also for this
observable, the $r=2$ scale-variation interval is always contained in the central plateau. In addition, we observe a progressive 
narrowing of distributions with increasing perturbative order. 

\begin{figure}[p]
	\centering
	\includegraphics[width=0.7\textwidth]{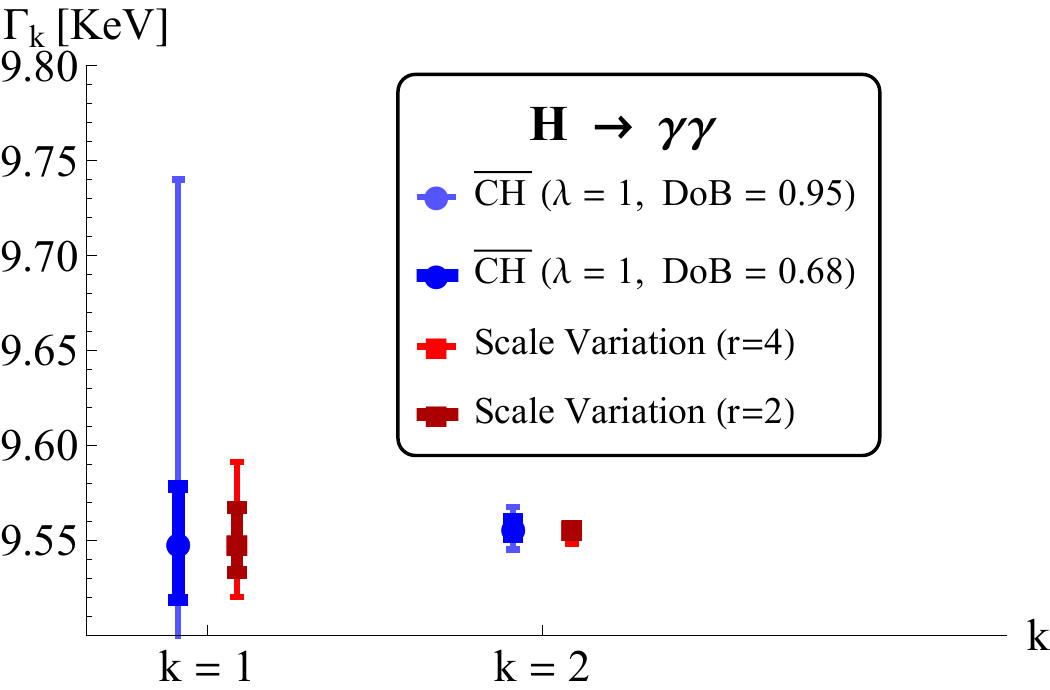}
	\caption{Size of the MHO uncertainty intervals at NLO and NNLO for the $H\to\gamma\gamma$ process with the \chbar\ model with $\lambda=1$, compared to those predicted by scale variation.}
	\label{fig:higgsphotonsbars}
\end{figure}

Finally we consider Higgs decay into two photons. In this case predictions at NLO are 
included within the LO uncertainty intervals determined using both the \chbar\ and the scale-variation ($r=2$) methods, as can be seen in 
Figure~\ref{fig:higgsphotonsbars}. On the other hand, we notice that in this case the $68\%$ DoB intervals are comparable in size with the 
intervals obtained from scale variation with $r=4$. This suggests that theoretical uncertainties of the Higgs decay into two photon process  determined with the scale-variation procedure using $r=2$ may be underestimated if one attempts to assign them a 68\% or larger heuristic CL.


\clearpage


\clearpage

\begin{figure}
	\centering	
	\includegraphics[width=0.9\textwidth]{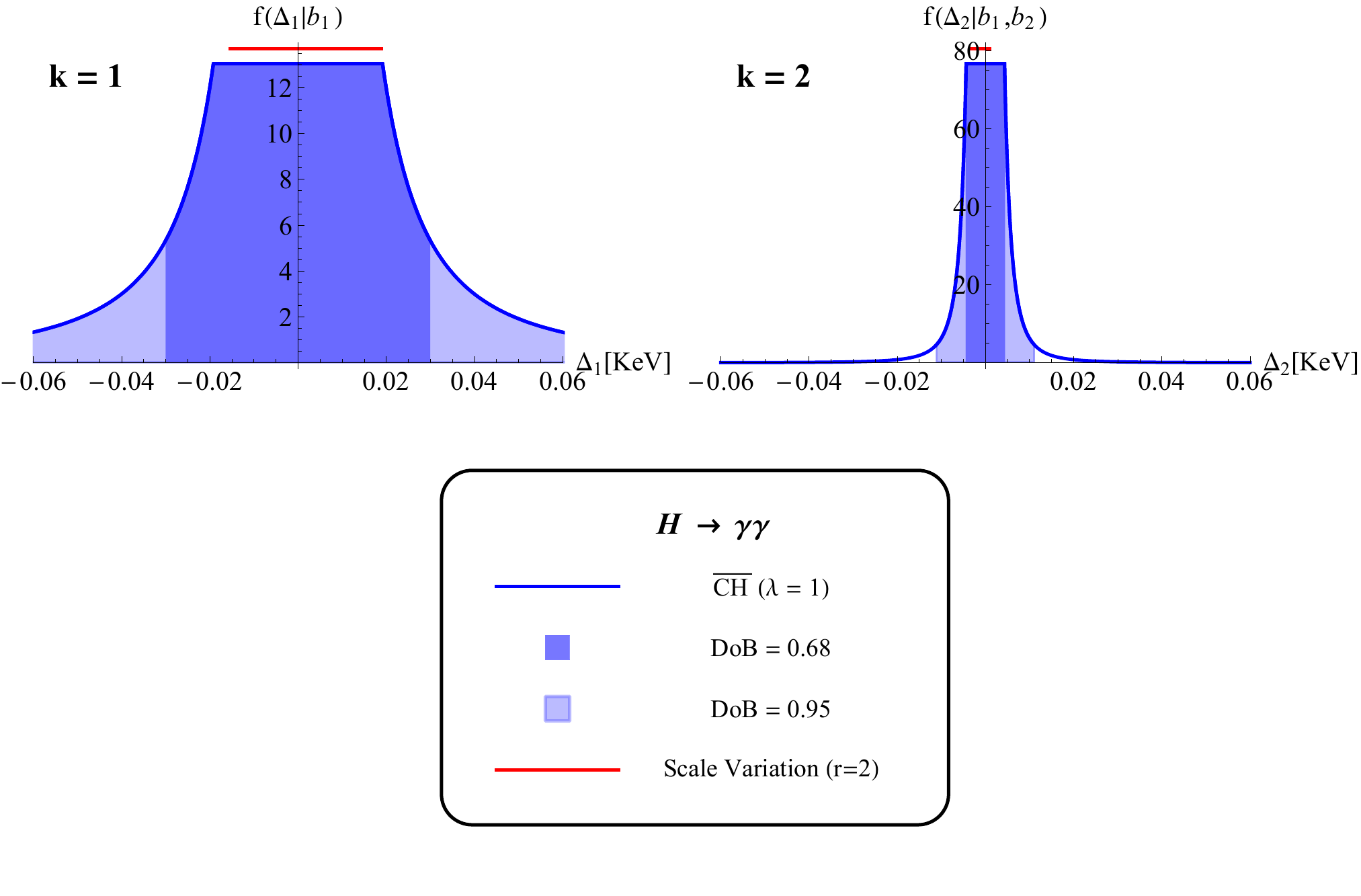}
	\caption{Posterior distribution for the remainder $\Delta_k$ (blue solid) for the $H\to \gamma\gamma$ process with the \chbar\ model, $68\%$ DoB interval (blue fill), $95\%$ DoB interval (light-blue fill), scale-variation interval with $r=2$ (red solid).}
	\label{fig:higgsphotonsposterior}
\end{figure}

\clearpage

\subsection{Processes with hadrons in the initial state}

We now consider a number of processes with hadrons in the initial state for which theoretical predictions are available 
at least up to NNLO, namely the production in proton-proton collisions of $Z$ and $W$ bosons, top-antitop pairs, and Higgs. These processes are either considered to be standard candles at hadronic colliders or are particularly 
relevant for LHC phenomenology. They provide precision tests of the Standard Model, and a significant discrepancy 
between their experimental measurement and theoretical predictions might be a hint of new physics at the TeV scale.

The numerical values of the perturbative coefficients together with the values used for the 
renormalisation and factorisation scales and the strong coupling constant, are collected in Table~\ref{tab:hadronic-obs} in Appendix~\ref{sec:tables}.
In Table~\ref{tab:hadr-cuts} in the same Appendix we list the cuts used in the computation of the observables.

In Table~\ref{tab:reshadr}, we summarise the results of our studies comparing, for each process, the size of the uncertainty 
intervals obtained with the \chbar\ method (68\% and 95\% DoB) to the ones obtained using the scale-variation procedure ($r=2$) 
at different perturbative orders.

\begin{figure}[p]
	\centering
	\includegraphics[width=0.7\textwidth]{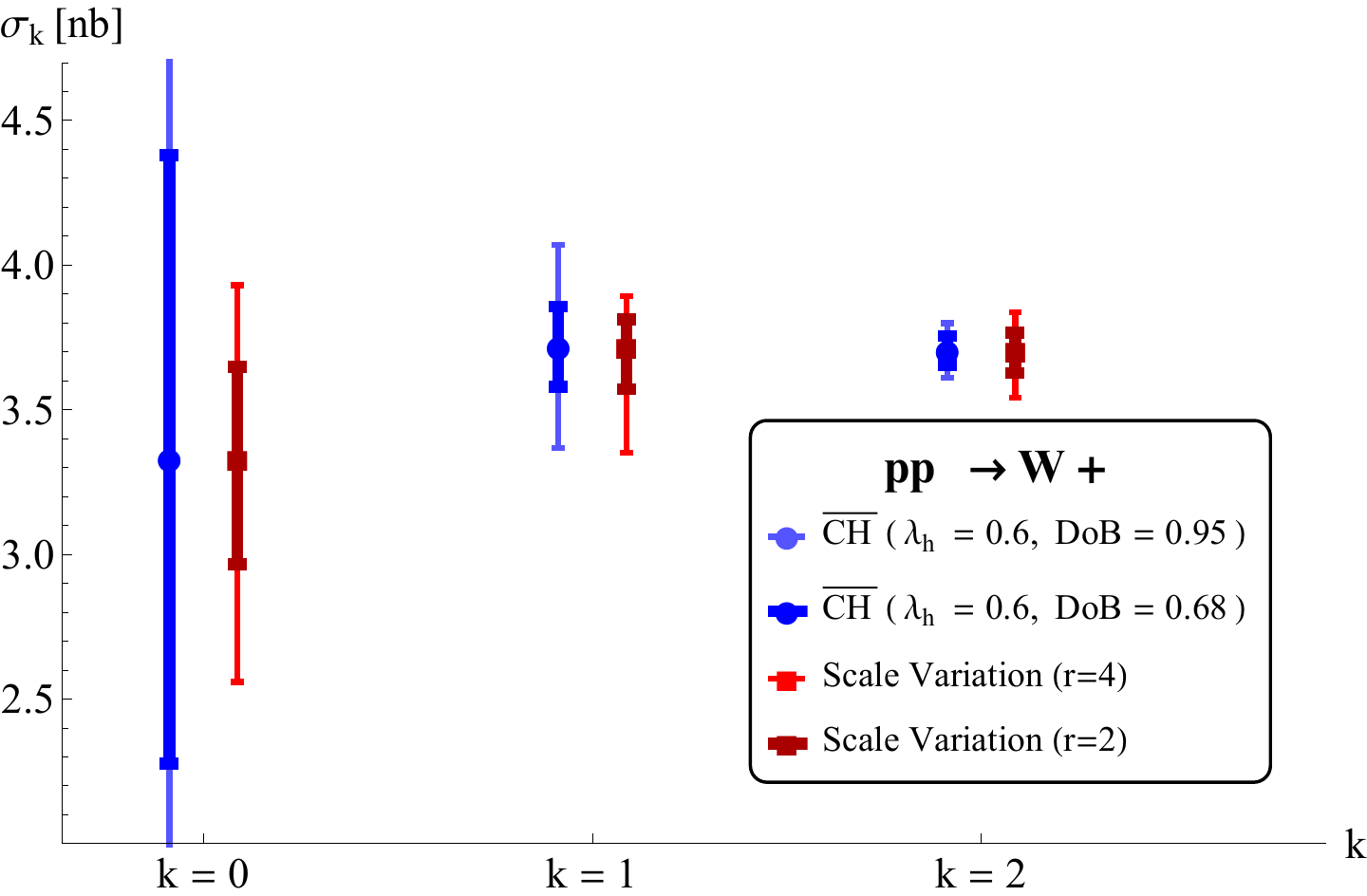}
	\caption{Size of the MHO uncertainty intervals at LO, NLO and NNLO for the $pp\to W^+$ process  at $\sqrt{S} = 8$~TeV  with the \chbar\ model with $\lambda_h=0.6$, compared to those predicted by scale
variation.}
	\label{fig:Wplusbars}
\end{figure}

\begin{figure}
	\centering
	\includegraphics[width=0.9\textwidth]{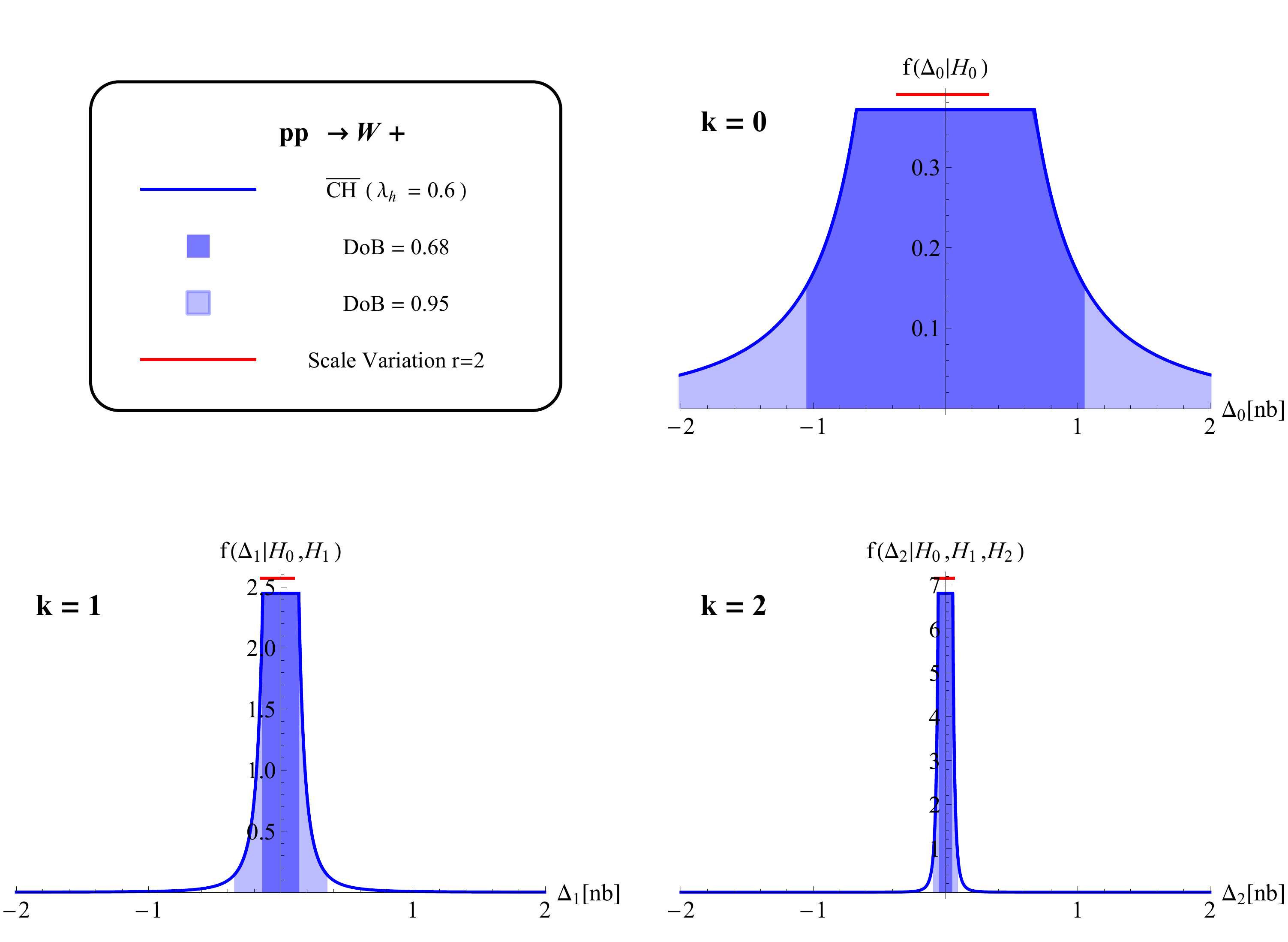}
	\caption{Posterior distribution for the remainder $\Delta_k$ (blue solid) for the $pp\to W^+$ process at $\sqrt{S} = 8$~TeV with the \chbar\ model, $68\%$ DoB interval (blue fill), $95\%$ DoB interval (light-blue fill), scale-variation interval (red solid).}
	\label{fig:Wplusposterior}
\end{figure}

\begin{figure}[p]
	\centering
	\includegraphics[width=0.7\textwidth]{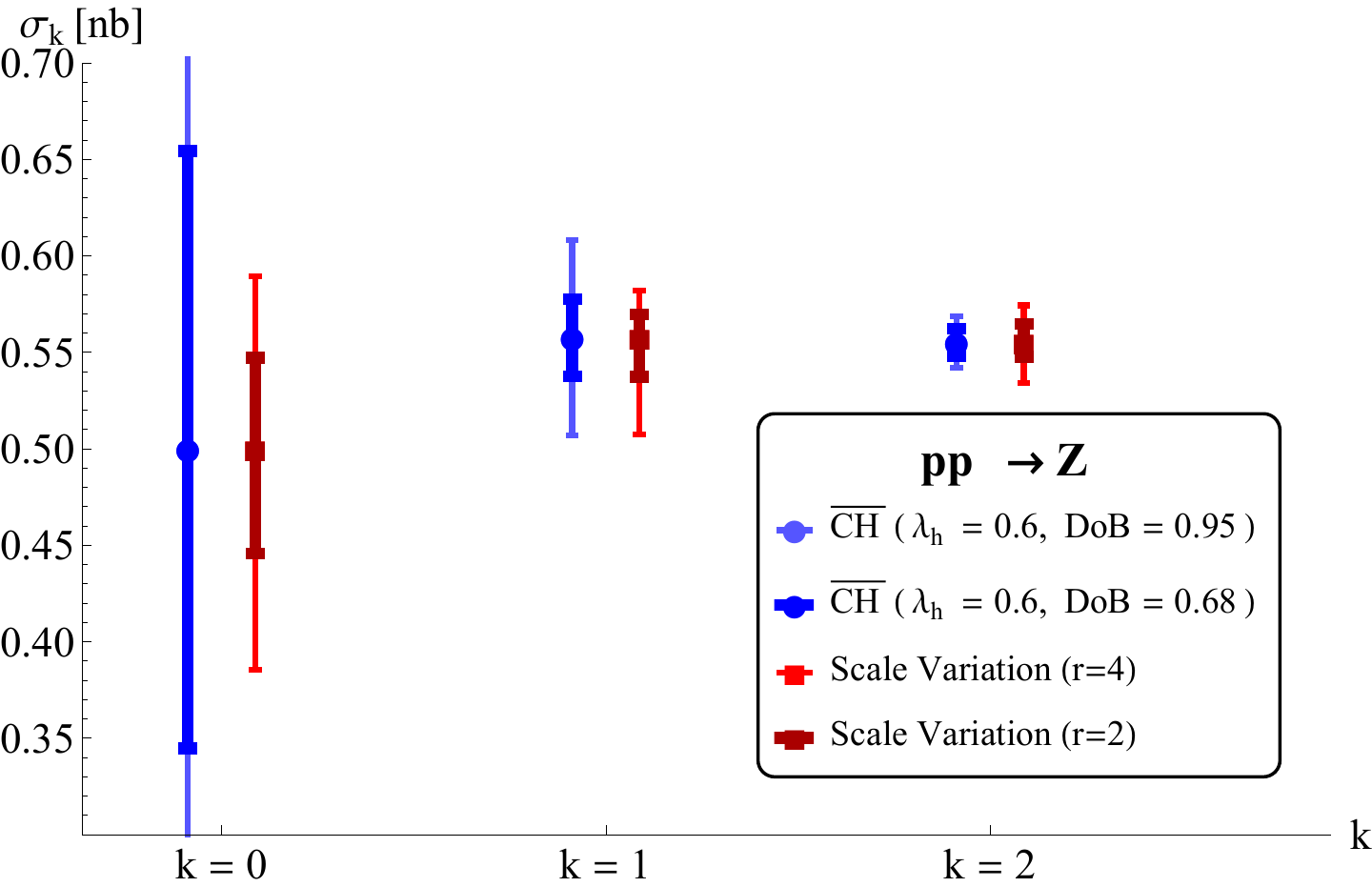}
	\caption{Size of the MHO uncertainty intervals at LO, NLO and NNLO for the $pp\to Z$ process at $\sqrt{S} = 8$~TeV with the \chbar\ model with $\lambda_h=0.6$, compared to those predicted by scale variation.}
	\label{fig:Zbars}
\end{figure}

\begin{figure}
	\centering
	\includegraphics[width=0.9\textwidth]{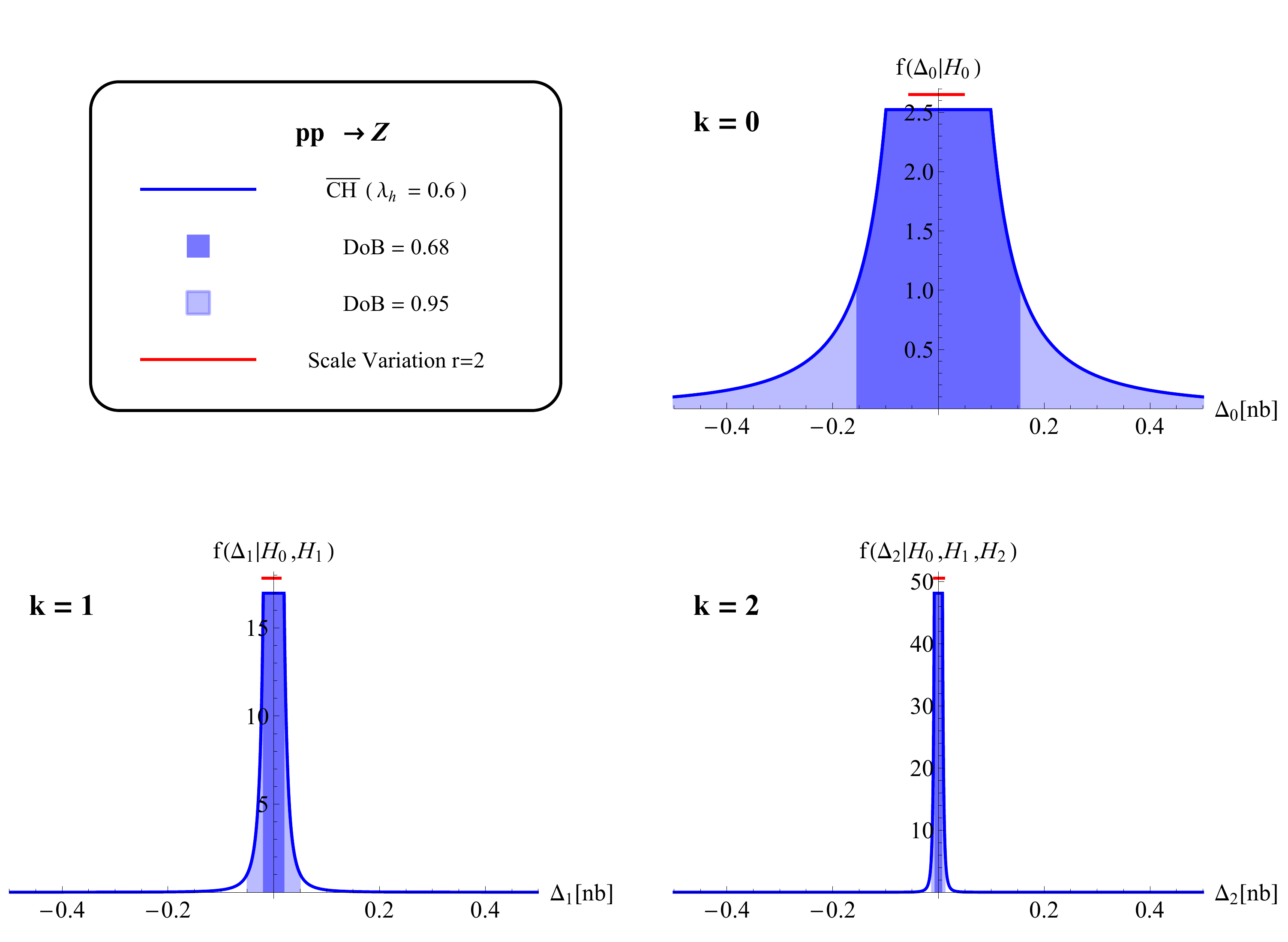}
	\caption{Posterior distribution for the remainder $\Delta_k$ (blue solid) for the $pp\to Z$ process at $\sqrt{S} = 8$~TeV with the \chbar\ model, $68\%$ DoB interval (blue fill), $95\%$ DoB interval (light-blue fill), scale-variation interval (red solid).}
	\label{fig:Zposterior}
\end{figure}

As far as weak vector boson production is concerned, a graphical comparison of predictions obtained with
the \chbar\ and the scale-variation methods is shown in Figures~\ref{fig:Wplusbars} and ~\ref{fig:Zbars}. We notice 
a similar behaviour for $W^+$ and $Z$ production. At LO the 68\% DoB uncertainty intervals obtained with 
the \chbar\ prescription are substantially larger than the intervals obtained from the scale-variation prescription with 
either $r=2$ or $r=4$.  At NLO the difference in size is reduced, while at NNLO the \chbar\ intervals turn out to be smaller
 in size than the scale-variation ones.
The posterior distributions for $\Delta_k$, shown in Figures~\ref{fig:Wplusposterior} and~\ref{fig:Zposterior}, show a progressive narrowing 
with the increase of the perturbative order.

\begin{figure}[p]
	\centering
	\includegraphics[width=0.7\textwidth]{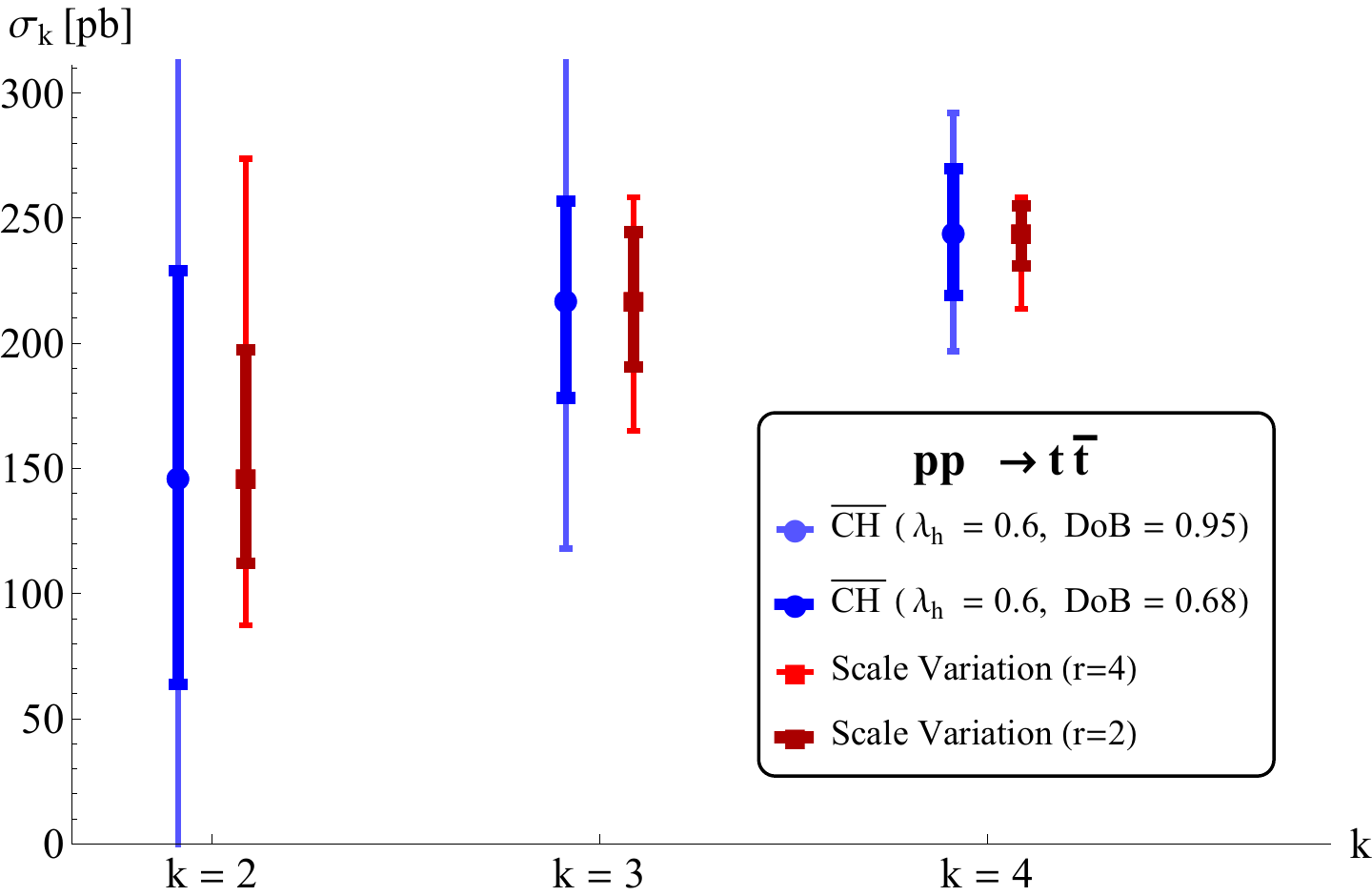}
	\caption{Size of the MHO uncertainty intervals at LO, NLO and NNLO for the $pp\to t\bar t $ process at $\sqrt{S} = 8$~TeV with the \chbar\ model with $\lambda_h=0.6$, compared to those predicted by scale variation.}
	\label{fig:ttbarbars}
\end{figure}

\begin{figure}
	\centering
	\includegraphics[width=0.9\textwidth]{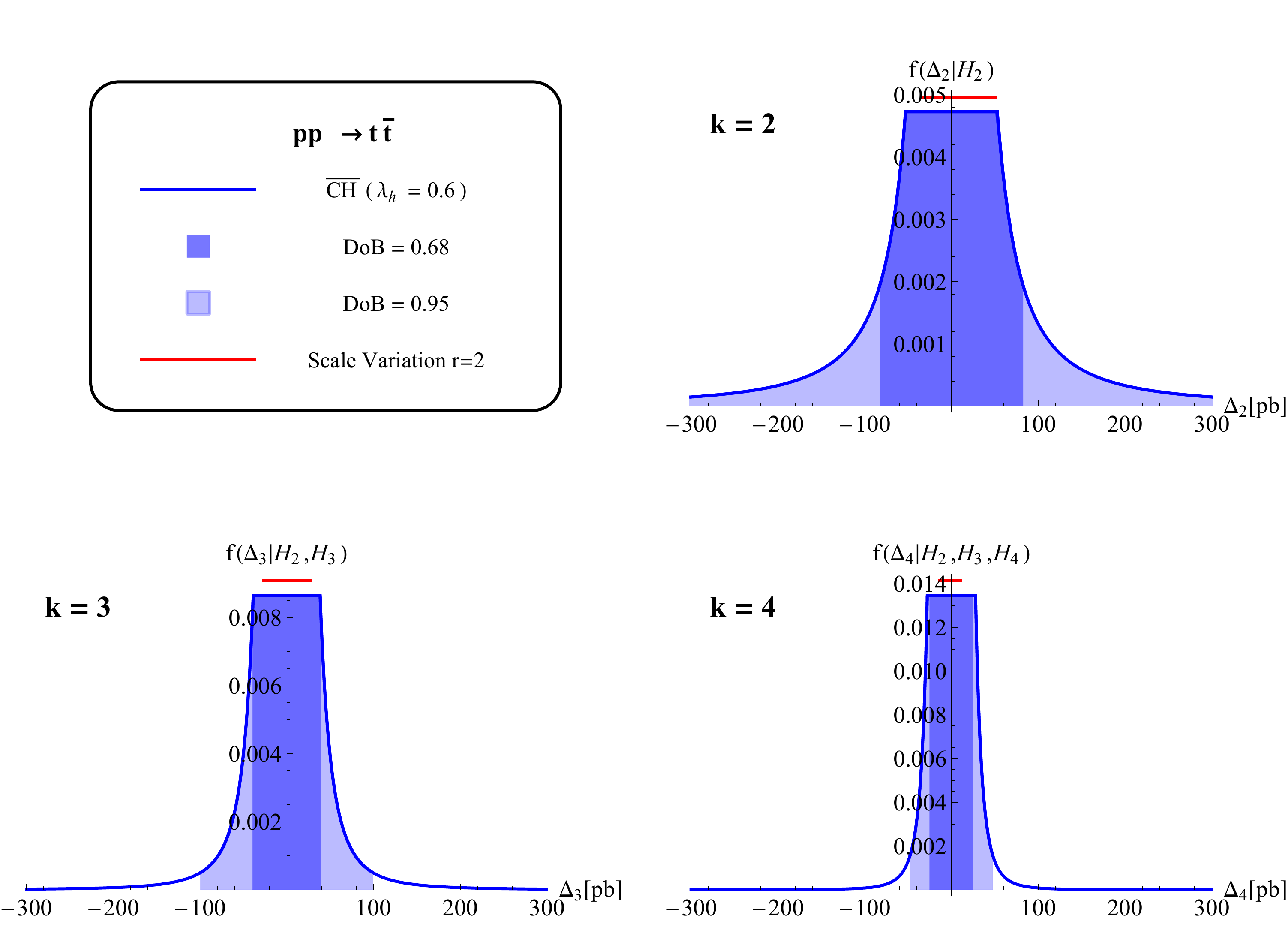}
	\caption{Posterior distribution for the remainder $\Delta_k$ (blue solid) for the $pp\to t\bar t$ process at $\sqrt{S} = 8$~TeV with the \chbar\ model, $68\%$ DoB interval (blue fill), $95\%$ DoB interval (light-blue fill), scale-variation interval (red solid).}
	\label{fig:ttbarposterior}
\end{figure}

For the top-pair production process the comparison of uncertainty intervals obtained using the \chbar\ and scale-variation methods
are shown in Figure~\ref{fig:ttbarbars}. In this case, the NLO result for the cross section is contained in the 
LO uncertainty band determined by $68\%$ DoB interval of the \chbar\ prescription, while this is not the case for the 
scale-variation interval obtained with $r=2$. On the other hand, the NNLO central prediction for the cross section
is outside the LO intervals computed with either method.
\chbar\ intervals with a DoB of 68\% are similar in size to the 
scale-variation intervals with $r=4$, and scale-variation intervals with $r=2$ are always smaller than the \chbar\ ones at 68\% DoB.
The posterior distributions for $\Delta_k$ plotted in Figure~\ref{fig:ttbarposterior} show the expected narrowing
as the perturbative order increases, and that the $r=2$ scale-variation interval is always contained in the flat part
of the distribution.

\begin{figure}[p]
	\centering
	\includegraphics[width=0.7\textwidth]{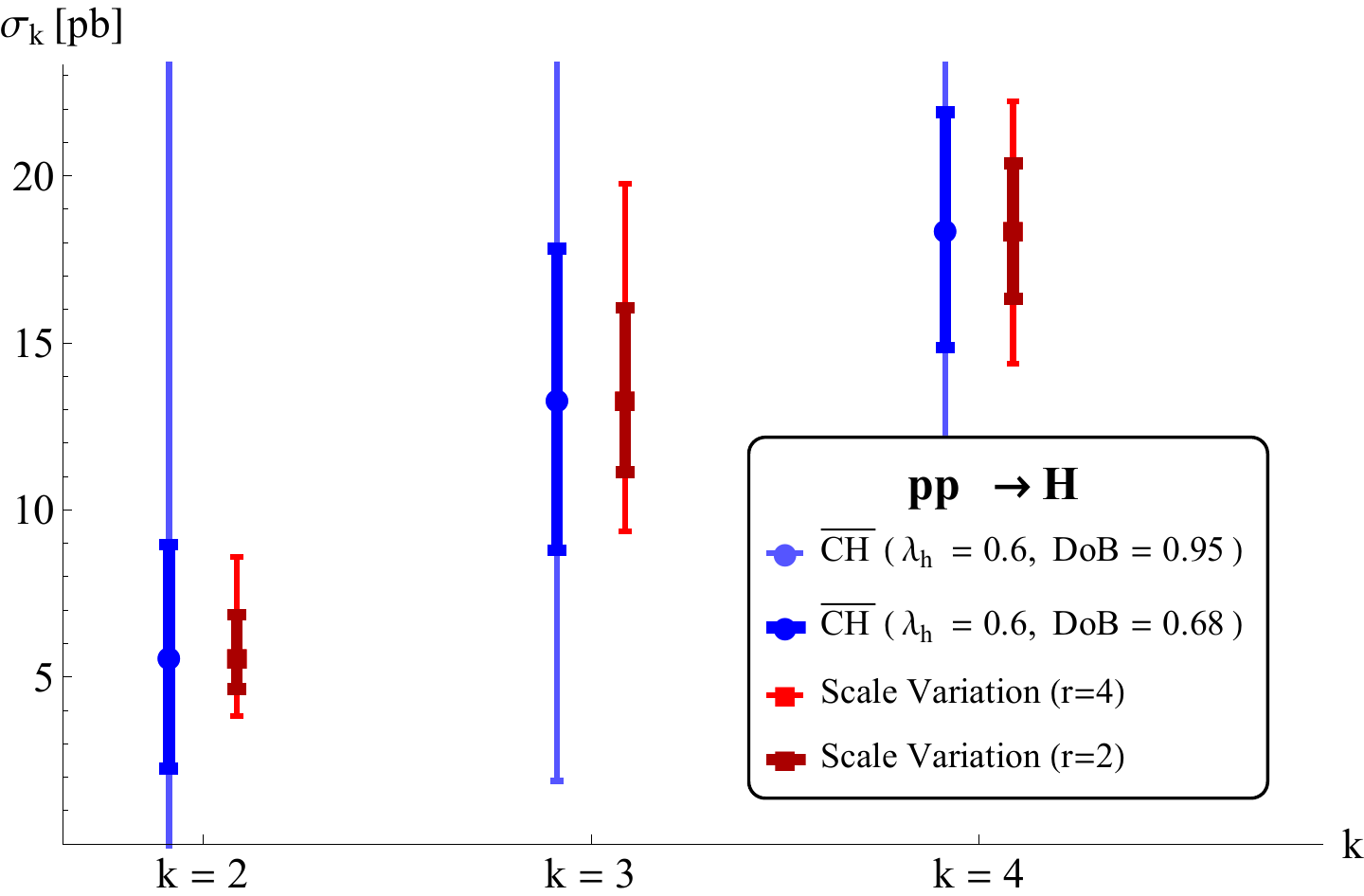}
	\caption{Size of the MHO uncertainty intervals at LO, NLO and NNLO for the $pp\to H$ via gluon fusion process at $\sqrt{S} = 8$~TeV with the \chbar\ model with $\lambda_h=0.6$, compared to those predicted by scale variation.}
	\label{fig:ppHbars}
\end{figure}

\begin{figure}
	\centering
	\includegraphics[width=0.9\textwidth]{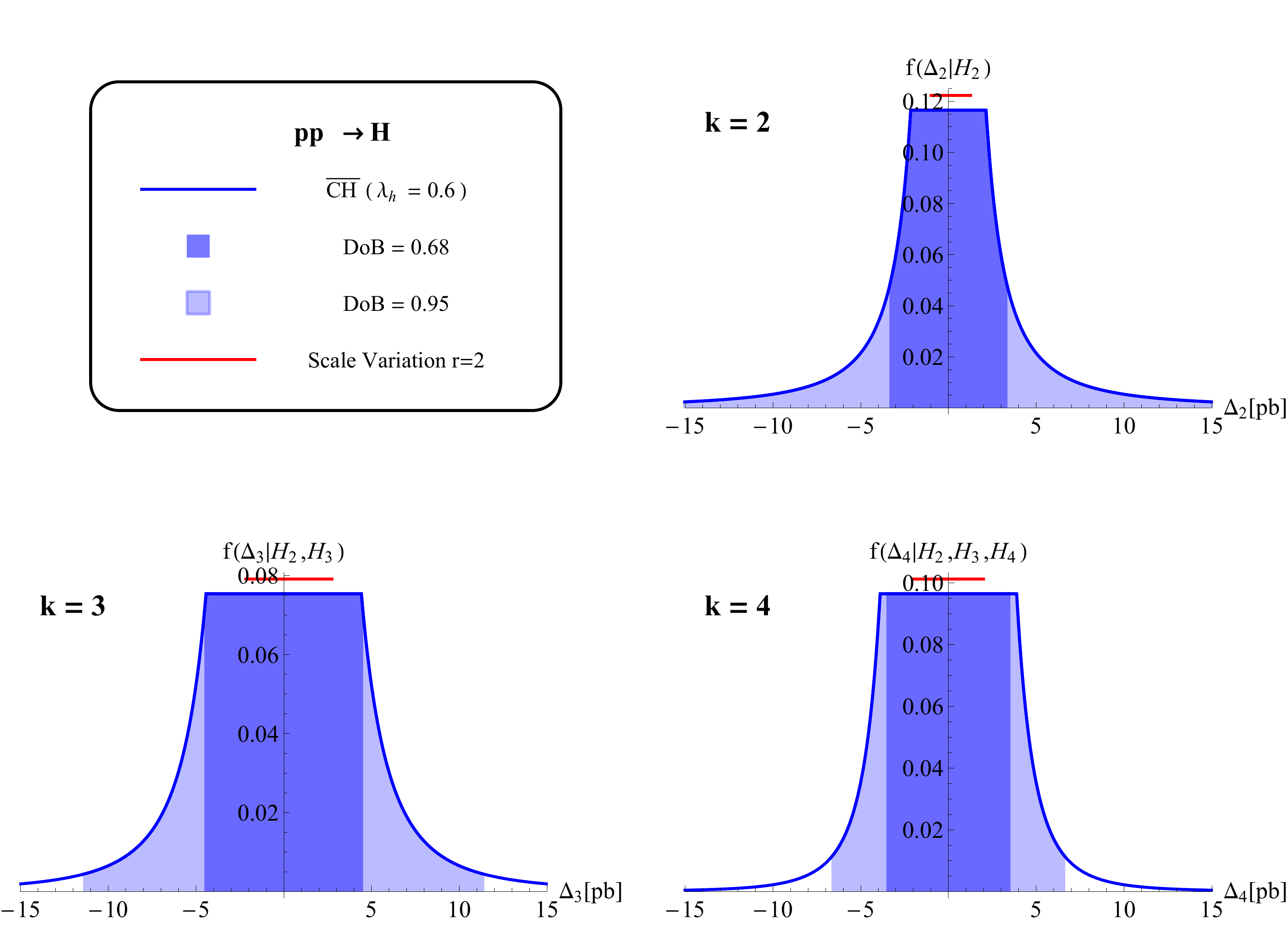}
	\caption{Posterior distribution for the remainder $\Delta_k$ (blue solid) for the $pp\to H$ via gluon fusion process at $\sqrt{S} = 8$~TeV with the \chbar\ model, $68\%$ DoB interval (blue fill), $95\%$ DoB interval (light-blue fill), scale-variation interval (red solid).}
	\label{fig:ppHposterior}
\end{figure}

As a final benchmark process, we consider Higgs production in proton-proton collisions, a process which is characterised
by large perturbative corrections. The graphical comparison in Figure~\ref{fig:ppHbars} shows that 
uncertainty intervals determined by scale variation with $r=2$ do not give a reliable error estimate for MHOUs for this 
observable, neither at LO nor at NLO. We observe that the \chbar\ 68\% DoB intervals are comparable in size with 
scale-variation intervals obtained with $r=4$. They both fail to properly estimate the large NLO correction, the central 
result at NLO being far outside the LO uncertainty band, and the error bars at NLO not even overlapping with the LO ones.
The \chbar\ method appears to produce a more reliable estimation of uncertainty intervals at higher orders, with the 68\% 
DoB intervals at NLO and NNLO showing a substantial overlap. 
The slowly-converging pattern of the perturbative expansion for the Higgs cross section prediction is reflected in the behaviour of the posterior 
distributions for $\Delta_k$, which are shown in Figure~\ref{fig:ppHposterior}. We observe that the narrowing of the posterior distribution 
with increasing perturbative order is now much less pronounced than for the other observables that we have considered in this Section. Also, the flat part of the posterior distribution for Higgs production broadens significantly when going from LO to NLO.

\clearpage

\begin{table}

\centering
\caption{Results for the analysis of missing higher order uncertainties for benchmark processes with initial-state hadrons. 
We quote the perturbative order $k$ at which the observable is calculated, the central value for the theoretical prediction at that order, the MHOs uncertainty 
intervals computed using the \chbar\ model at 68\% DoB and 95\% DoB, and the uncertainty interval obtained using  the scale variation (SV)  procedure with $r=2$.}
  \label{tab:reshadr}
  
\subtable[MHOUs for $W$ production in the Drell-Yan process at $\sqrt{S} = 8$~TeV. The perturbative series of the observable 
at the order $k$ is defined as $\sigma_k(pp \rightarrow W^+ \rightarrow l^{+} \nu_l) = \sum_{n=0}^k \alpha_s^n h_n$.]
{
  \centering
  \ra{1.1}
  \begin{tabular}{@{}lcccc@{}}\toprule
    \multicolumn{5}{c}{\text{\LARGE $pp \rightarrow W^+ \rightarrow l^{+} \nu_l$ }} \\
    \toprule
    Order &  $\sigma_k$[nb] & \chbar$_\mathrm{68\% DoB}$ & \chbar$_\mathrm{95\% DoB}$ & $\mathrm{SV}_{r=2}$ \\ \midrule
    \vspace{.2cm}
    $k = 0$ & 3.328 & $\pm 1.051$ & $\pm 6.729$ & {\Large $\substack{+0.319\\-0.362}$} \\
    \vspace{.2cm}
    $k = 1$ & 3.718 & $\pm 0.139$ & $\pm 0.351$ & {\Large $\substack{+0.095\\-0.147}$} \\
    \vspace{.2cm}
    $k = 2$ & 3.704 & $\pm 0.050$ & $\pm 0.094$ & {\Large $\substack{+0.061\\-0.077}$} \\
    \bottomrule
  \end{tabular}
  \label{tab:DYW}
}

\subtable[MHOUs for $Z$ production in the Drell-Yan process at $\sqrt{S} = 8$~TeV. The perturbative series of the 
observable at the order $k$ is defined as $\sigma_k(pp \rightarrow Z \rightarrow e^+e^-) = \sum_{n=0}^k \alpha_s^n h_n$.]{
  \centering
  \ra{1.1}
  \begin{tabular}{@{}lcccc@{}}\toprule
    \multicolumn{5}{c}{\text{\LARGE $pp \rightarrow Z \rightarrow e^+e^-$ }} \\
    \toprule
    Order &  $\sigma_k$[nb] & \chbar$_\mathrm{68\% DoB}$ & \chbar$_\mathrm{95\% DoB}$ & $\mathrm{SV}_{r=2}$ \\ \midrule
    \vspace{.2cm}
    $k = 0$ & 0.4995 & $\pm 0.1548$ & $\pm 0.9907$ & {\Large $\substack{+0.047\\-0.054}$} \\
    \vspace{.2cm}
    $k = 1$ & 0.5574 & $\pm 0.0201$ & $\pm 0.0507$ & {\Large $\substack{+0.012\\-0.020}$} \\
    \vspace{.2cm}
    $k = 2$ & 0.5551 & $\pm 0.0071$ & $\pm 0.0133$ & {\Large $\substack{+0.010\\-0.007}$} \\
    \bottomrule
  \end{tabular}
  \label{tab:DYZ}

}

\subtable[MHOUs for $t\bar{t}$ production at $\sqrt{S} = 8$~TeV. The perturbative series of the observable at the 
order $k$ is defined as $\sigma_k(pp \rightarrow t\bar{t}) = \sum_{n=2}^k \alpha_s^n h_n$.]
{
  \centering
  \ra{1.1}
  \begin{tabular}{@{}lcccc@{}}\toprule
    \multicolumn{5}{c}{\text{\LARGE $pp \rightarrow t\bar{t}$ }} \\
    \toprule
    Order &  $\sigma_k$[pb] & \chbar$_\mathrm{68\% DoB}$ & \chbar$_\mathrm{95\% DoB}$ & $\mathrm{SV}_{r=2}$ \\ \midrule
    \vspace{.2cm}
    $k = 2$ & 146.32 & $\pm 82.61$ & $\pm 528.76$ & {\Large $\substack{+51.08\\-34.32}$} \\
    \vspace{.2cm}
    $k = 3$ & 217.38 & $\pm 39.32$ &  $\pm 99.46$ & {\Large $\substack{+26.94\\-26.89}$} \\
    \vspace{.2cm}
    $k = 4$ & 244.36 & $\pm 25.24$ &  $\pm 47.60$ & {\Large $\substack{+12.42\\-13.52}$} \\
    \bottomrule
  \end{tabular}
  \label{tab:ttbar}
}

\subtable[MHOUs for Higgs production in gluon fusion at $\sqrt{S} = 8$~TeV. The perturbative series of the observable 
at the order $k$ is defined as $\sigma_k(pp \rightarrow H) = \sum_{n=2}^k \alpha_s^n h_n$.]
{
  \centering
  \ra{1.1}
  \begin{tabular}{@{}lcccc@{}}\toprule
    \multicolumn{5}{c}{\text{\LARGE $pp \rightarrow H$ }} \\
    \toprule
    Order &  $\sigma_k$[pb] & \chbar$_\mathrm{68\% DoB}$ & \chbar$_\mathrm{95\% DoB}$ & $\mathrm{SV}_{r=2}$ \\ \midrule
    \vspace{.2cm}
    $k = 2$ & 5.6 & $\pm 3.35$ & $\pm 21.46$ & {\Large $\substack{+1.26\\-0.98}$} \\
    \vspace{.2cm}
    $k = 3$ & 13.3 & $\pm 4.51$ & $\pm 11.42$ & {\Large $\substack{+2.74\\-2.17}$} \\
    \vspace{.2cm}
    $k = 4$ & 18.37 & $\pm 3.52$ & $\pm 6.65$ & {\Large $\substack{+2.00\\-2.06}$} \\
    \bottomrule
  \end{tabular}
  \label{tab:ppH}
}

\end{table}

\clearpage


\clearpage


\clearpage

\vspace{2cm}

\clearpage


\clearpage

\section{Conclusions and outlook}

In this paper we have investigated the performance of two approaches in estimating the theoretical uncertainties due to missing higher orders in perturbative QCD computations.
The first one is the widely used prescription of varying the unphysical factorisation and renormalisation 
scales around a central value. The second (\chbar) is a modified version of the Bayesian approach 
introduced by Cacciari and Houdeau in~\cite{Cacciari:2011ze}.

We have performed a global survey based on a wide set of perturbative observables. Within this set we have  considered two categories, characterised by the absence (``non-hadronic'') or the presence (``hadronic'') of hadrons in the initial state of a process, and we have analysed them separately. The outcome of this survey has allowed us to assign an heuristic confidence level to the uncertainty 
intervals returned by the scale-variation approach and, in a separate analysis, to determine an optimal 
expansion parameter to be employed in the \chbar\ Bayesian approach.

We have found that, in the scale-variation approach, the standard variation within a factor of two with respect to 
the central scale can lead to uncertainty intervals whose heuristic confidence level (CL) falls short of a conventional 68\%, thereby leading to possibly underestimate the real uncertainty.
 This is true for both the non-hadronic observables 
and, more markedly, for the hadronic ones. We have determined that the rescaling factor needed to obtain 
 68\%-heuristic CL intervals is usually larger than two, with the specific value depending on the class of observables under consideration 
and on the prescription used. In general, and conservatively, a  rescaling factor of between 
three and four appears more likely to give an estimation of the missing higher orders uncertainty that is consistent with a 68\%-heuristic CL interpretation of the 
scale-variation intervals.

Our analysis of the \chbar\ approach has allowed us to determine that $\alpha_s$ is an appropriate expansion parameter for non-hadronic observables, while a slightly larger parameter, $\alpha_s/\lambda_h$, with $\lambda_h \simeq 0.6$, appears to be more appropriate for hadronic observables.

Armed with the determination of these expansion parameters from the global survey,  we have then  compared the performances of the scale-variation and the \chbar\ Bayesian approaches in the estimation of the 
MHOUs for a set of benchmark observables of particular interest for LHC physics, namely 
the production in proton-proton collisions of electroweak vector bosons, top-antitop pairs and Higgs.
The two approaches perform similarly in estimating, to a given heuristic confidence level for the scale-variation approach or to a given credibility level for the Bayesian \chbar\ approach, the MHOUs for these observables, provided that a rescaling factor larger than two is used for scale variation. 
More importantly, however, the \chbar\ approach additionally provides a full probability density distribution for the missing higher orders uncertainty. This probability density can then be used to combine in a meaningful way the MHOU with uncertainties of different origin, e.g. experimental ones.

We conclude by commenting on two possible avenues for further development. First, in this paper we have determined the optimal expansion parameter for the perturbative series in the \chbar\ approach by performing a frequentist analysis of a set of known observables. One could envisage replacing this analysis with an additional prior on the expansion parameter. Second, in this work we have adopted, in the form of the \chbar\ model, an as generic Bayesian approach as possible, meant to be applied to a wide and open-ended class of perturbative observables. One could instead envisage developing other, more refined Bayesian models, crafted to work on a more restricted and more uniform class of observables. In such a case, more specific knowledge about the perturbative behaviour of the observables would be input into the model, and a trade-off would exist between this amount of information and the model's eventual predictivity. We leave exploration of these avenues to further work.

\section*{Acknowledgements}

This work was supported by in part by the ERC advanced grant Higgs@LHC, by the French
Agence Nationale de la Recherche, under grant ANR-10-CEXC-009-01, 
by the EU ITN grant LHCPhenoNet, PITN-GA-2010-264564 
and by the ILP LABEX (ANR-10-LABX-63) supported by French state funds
managed by the ANR within the Investissements d'Avenir programme under
reference ANR-11-IDEX-0004-02.

\appendix

\section{Numerical values of perturbative coefficients}
\label{sec:tables}
The observables used in our analyses are listed in Table~\ref{tab:non-hadronic-obs} and Table~\ref{tab:hadronic-obs}.

Table~\ref{tab:non-hadronic-obs}
gives the perturbative coefficients for non-hadronic observables, i.e. without hadrons in the initial state. They are given in the form
\begin{align}
  O_k(Q,\mu_r) = \sum_{n=l}^k \alpha_s^n(\mu_r) c_n(Q,\mu_r) \, ,
\end{align}
where $\mu_r$ is the renormalisation scale of the strong coupling, which we choose equal to the typical scale of the process, $Q$, when calculating the coefficients given in the table. 
We recall that for this class of observables, the coefficients $c_0$ are not used in the Bayesian analysis.

Hadronic observables (i.e. with hadrons in the initial state) are written as
\begin{align}
  O_k(Q,\mu_r,\mu_f) =\sum_{n=l}^k \alpha_s^n(\mu_r) h_n(\mu_r, \mu_f) \equiv \sum_{n=l}^k \alpha_s^n(\mu_r) {\cal L}(\mu_f) \otimes C_n(Q,\mu_r,\mu_f) \, .
\end{align}
We present in Table~\ref{tab:hadronic-obs}, setting $\mu_r = \mu_f  = Q$, the coefficients obtained after 
convolution with the parton-parton flux, as explained in Section~\ref{sec:had-observables}. Finally, in Table~\ref{tab:hadr-cuts} 
we list the cuts applied in the numerical computations of hadronic observables.

\begin{table}[h]
\small
\centering
\ra{1.1}
\begin{tabular}{@{}lrrrrrrrrr@{}}\toprule
\multicolumn{10}{c}{\text{\LARGE Non-Hadronic observables }} \\
\multicolumn{5}{c}{Parameters} & \multicolumn{5}{c}{Coefficients} \\
\cmidrule{1-4} \cmidrule{6-10} Observable & $Q (\mathrm{GeV})$ & $\alpha_s(Q)$ & $l$ && $c_l$ & $c_{l+1}$ & $c_{l+2}$ & $c_{l+3}$ & $c_{l+4}$ \\
\cmidrule{1-4} \cmidrule{6-10}
$R=\frac{\sigma(e^+e^-\to\text{hadr})}{\sigma_{0}(e^+e^-\to\text{hadr})}$ & 91.19 & 0.118 & 0 && 1 & 0.318 & 0.143 & -0.413 &  \\
Bjorken sum rule &  91.19 & 0.118 & 0 && 1 & -0.212 & -0.238 & -0.274 &  \\
GLS sum rule & 91.19 & 0.118 & 0 && 6. & -1.910 & -1.773 &-1.117 & \\
$\frac{\Gamma(b\to c e\bar{\nu}_{e})}{\Gamma_0(b\to c e\bar{\nu}_{e})}$ & 4.6 & 0.22 & 0 && 1 & -0.566 & -1.408 & &\\
$\Gamma(Z\to\text{hadr})$[GeV] & 91.19 & 0.118 & 0 && 1.674 & 0.533 & 0.130 &  -0.837 & -1.173  \\
$\frac{\Gamma(Z\to b\bar{b})}{\Gamma_0(Z\to b\bar{b})|_{m_b=0}}$ & 91.19 & 0.118 & 0 && 0.997 & 0.315 & -0.156 & -0.796  & \\
$\Gamma(H\to gg)$ [MeV] & 125 & 0.113 & 2 && 14.43 & 82.28  & 223.6 & 181.6 & \\
$\Gamma(H\to b\bar{b})|_{m_b=0}$[MeV] & 125 & 0.113 & 0 && 1.850 & 3.338 & 5.465 & 2.492 & -15.685 \\
$\Gamma(H\to\gamma\gamma)$ [KeV] & 125 & 0.113 & 0 && 9.379 & 1.494 & 0.627 & &\\
$\langle$3-jets Thrust$\rangle$ &  91.19 & 0.118 & 1 && 0.030 & 0.149 & 0.686 & &\\
$\langle$3-jets Heavy jet mass$\rangle$ & 91.19 & 0.118 & 1 && 0.030 & 0.069 & 0.141 & & \\
$\langle$3-jets Wide jet broadening$\rangle$ &  91.19 & 0.118 & 1 && 0.054 & 0.098 & 0.166 & & \\
$\langle$3-jets Total jet broadening$\rangle$ & 91.19 & 0.118 & 1 && 0.054 & 0.356 & 1.219 & &\\
$\langle$3-jets C parameter$\rangle$ & 91.19 & 0.118 & 1 && 0.387 & 1.933 & 8.731 & &\\
$\langle$3-to-2 jet transition$\rangle$ & 91.19 & 0.118 & 1 && 0.013 & 0.029 & 0.044 & &\\
$\gamma_{ns}^{(+)}(N=2)$ & 91.19 & 0.118 & 1 && 0.283 & 0.206 & 0.081  & &\\
$\gamma_{qq}(N=2)$ & 91.19 & 0.118 & 1 && 0.283 & 0.143 & -0.068 & & \\
$\gamma_{qg}(N=2)$ & 91.19 & 0.118 & 1 && -0.265 & -0.239 & 0.058 & & \\
\bottomrule
\end{tabular}
\caption{QCD perturbative corrections for observables without initial-state hadrons. The coefficients $c_n$ are defined by $O_k=\sum_{n=l}^k \alpha_s^n c_n$, where $O_k$ is an observable computed at $k^\mathrm{th}$ order in perturbative QCD.
         Notice that $l=0$ coefficients are not used in the Bayesian non-hadronic analysis.}
\label{tab:non-hadronic-obs}
\end{table}

\begin{table}[h]
  \small
\centering
\ra{1.1}
\begin{tabular}{@{}lrrrrrrrr@{}}\toprule
\multicolumn{9}{c}{\text{\LARGE Hadronic observables }} \\
\multicolumn{4}{c}{Parameters} & \multicolumn{4}{c}{Coefficients} \\
\cmidrule{1-4} \cmidrule{6-9} Observable (LHC, $\sqrt{S}=8$ TeV) & $Q$ & $\alpha_s(Q)$ & $l$ && $h_l$ & $h_{l+1}$ & $h_{l+2}$ & \\ 
\cmidrule{1-4} \cmidrule{6-9}
$\sigma(pp\to H)$ [pb] & 125 & 0.115  & 2 && 424. & 5072. &  29097. \\
$\sigma(pp \to b\bar{b}\to H)$ 5FS [pb]& 125 & 0.113 & 0 && 0.402 & -0.854 & -4.951 \\
$\sigma(pp\to Z^*+X\to ZH+X)$ [pb]& 216.2 & 0.105 & 0 && 0.332 & 0.587 & 2.734 \\
$\sigma(pp\to W^*+X\to WH+X)$ [pb]& 205.6 & 0.105 & 0 && 0.626 & 1.108 & 1.834 \\
$\sigma(pp\to b\bar{b})$ [$\mu$b]& 20 & 0.155 & 2 && 5371. & 31190. & \\
$\sigma(pp\to t\bar{t})$ [pb]& 173.3 & 0.108 & 2 && 12449. & 55769. & 195299. \\
$\sigma(pp\to Z+X\to e^+e^-$ [nb]& 91.19 & 0.119 & 0 && 0.500 & 0.486 & -0.164 \\
$\sigma(pp\to Z+j)$ [nb]& 91.19 & 0.119 & 1  && 1.186 & 2.831 &  \\
$\sigma(pp\to Z+2j)$ [nb]& 91.19 & 0.119 & 2  && 3.659 & 5.138 &  \\
$\sigma(pp\to ZZ)$ [fb]& 182.4 & 0.108 & 0 && 4.949 &  14.311 & \\
$\sigma(pp\to W^-+X\to e^-+\nu_e+X)$ [nb] & 80.4 & 0.121 & 0 && 2.241 & 2.108 & -2.074 \\
$\sigma(pp\to W^++X\to e^++\bar{\nu}_e+X)$ [nb] & 80.4 & 0.121 & 0 && 3.328 & 3.212 & -0.922  \\
$\sigma(pp\to W^++j)$ [nb]& 80.4 & 0.121 & 1 && 6.182 & 17.547 &  \\
$\sigma(pp\to W^-+j)$ [nb]& 80.4 & 0.121 & 1 && 4.385 & 11.573 &  \\
$\sigma(pp\to W^++2j)$ [nb]& 80.4 & 0.121 & 2 && 19.450 & 28.868 &  \\
$\sigma(pp\to W^-+2j)$ [nb]& 80.4 & 0.121 & 2 && 12.993 & 20.632 &  \\
$\sigma(pp\to W^+W^-)$ [pb]& 160.8 & 0.109 & 0 && 0.175 & 0.742 &  \\
\bottomrule
\end{tabular}
\caption{QCD perturbative corrections for observables with initial-state hadrons. The coefficients $h_n$ are defined by $O_k=\sum_{n=l}^k \alpha_s^n h_n \equiv \sum_{n=l}^k \alpha_s^n {\cal L}\otimes C_n$, where $O_k$ is an observable computed at $k^\mathrm{th}$ order in perturbative QCD.
All observables have been computed for the LHC (proton-proton collisions at $\sqrt{S}=8$ TeV) with the cuts given in Table~\ref{tab:hadr-cuts}.}
\label{tab:hadronic-obs}
\end{table}

\begin{table}[h]
  \small
\centering
\ra{1.1}
\begin{tabular}{@{}cc@{}}\toprule
\multicolumn{2}{c}{\text{\LARGE Hadronic analysis cuts}} \\
Cut & Description \\
\midrule
 $ 0 \leq m_{34}^{\text{min}} \leq 14$~TeV & Invariant mass of particles $3$ and $4$ in the process\\
 $ 0 \leq m_{34}^{\text{min}} \leq 14$~TeV & Invariant mass of particles $5$ and $6$ in the process\\
 $ m^T_{34} \geq 0 $ GeV & Transverse mass of particles $3$ and $4$ in the process\\
 Anti-$k_T$ \cite{Cacciari:2008gp}, $R=0.4$ & Jet algorithm \\
 $p_T^{\text{jet}} \geq 15$ GeV & Jet transverse momentum $p_T$\\
 $0 \leq |\eta^{\text{jet}}| \leq 3.5$ & Jet pseudorapidity\\
 $p_T^{\text{lept}} \geq 20$ GeV & Lepton transverse momentum\\
 $0 \leq |\eta^{\text{lept}}| \leq 2.5$ & Lepton pseudorapidity\\
 $p_T^{\text{miss}} \geq 25$ GeV & Missing (neutrinos) transverse momentum\\
 $\Delta R_{jj} > 0.5 $ & Jet-jet separation $\Delta R_{jj} = \sqrt{\Delta\eta^2_{jj} + \Delta\phi^2_{jj}}$\\
 $\Delta R_{jl} > 0.4$ & Jet-lepton $\Delta R_{jl} = \sqrt{\Delta\eta^2_{jl} + \Delta\phi^2_{jl}}$\\
 $\Delta R_{ll} > 0.4$ & Lepton-lepton $\Delta R_{ll} = \sqrt{\Delta\eta^2_{ll} + \Delta\phi^2_{ll}}$\\
 $\Delta \eta_{jj} > 0$ & Jet-jet pseudorapidity separation \\
\bottomrule
\end{tabular}
\caption{Cuts used in the hadronic analysis. }
\label{tab:hadr-cuts}
\end{table}

\clearpage

\section{Hadronic observables: convoluted coefficients vs dominant Mellin moment method}
\label{section:mellinAppendix}
Our preferred extension of the \chbar\ model to hadronic observables is the one based on the convoluted coefficients (see
Section~\ref{sec:had-observables}), due to its 
ability to capture effectively the full complexity of a process with initial-state hadrons and  multiple partonic channels.  However, it is instructive to 
compare its results with those of the Mellin-moment approach described in Section~\ref{sec:had-observables}. In this Appendix we review two processes: Higgs production in 
gluon fusion, where, as we will see, the Mellin method and the coefficient one return equivalent results, and the Drell-Yan process, 
where the Mellin method fails to capture the essence of the process.

\subsection{Higgs production in gluon fusion at the LHC}

\begin{figure}
  \centering
  \includegraphics[width=0.7\textwidth]{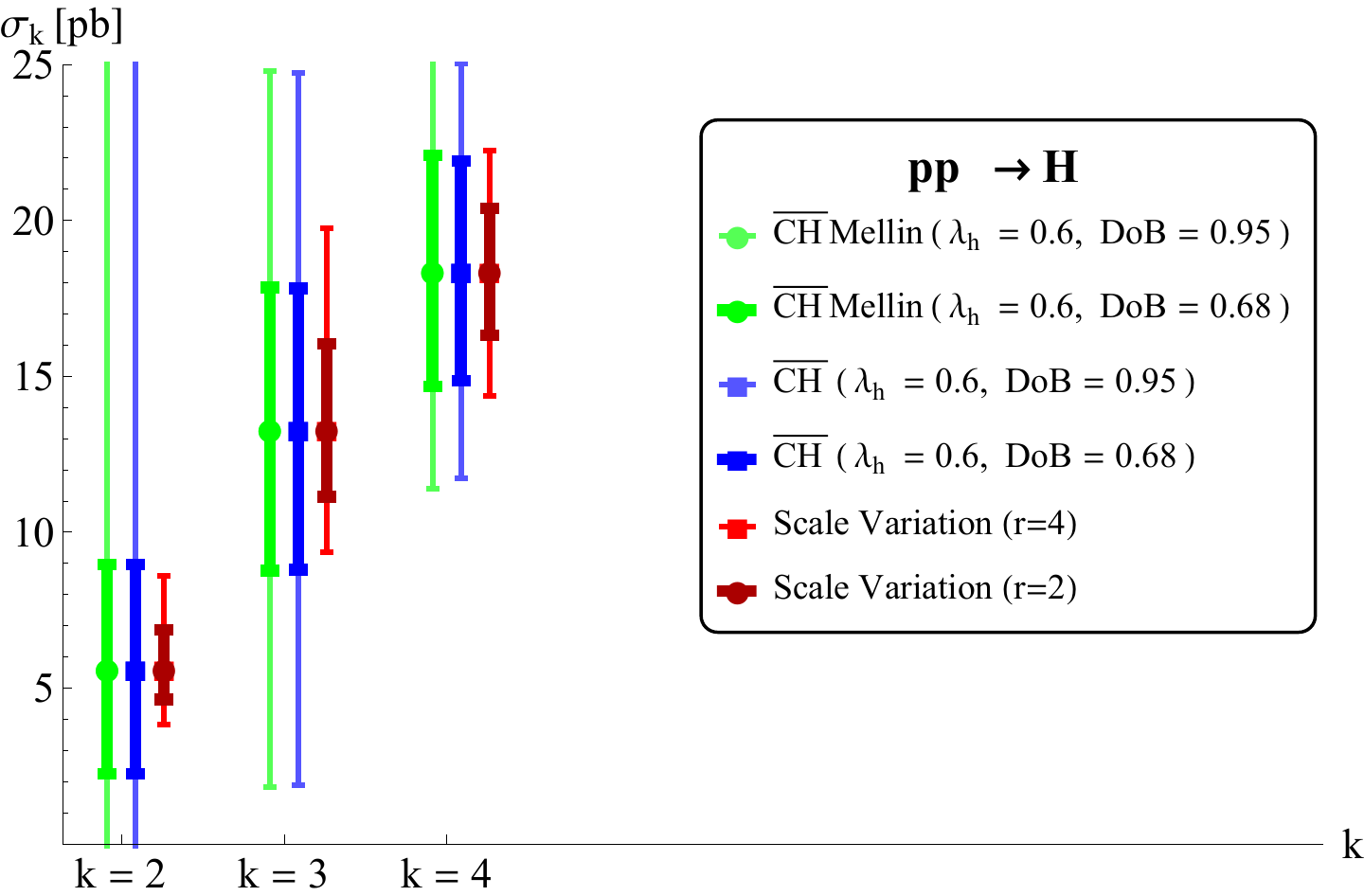}
	\caption{Size of the MHO uncertainty intervals at LO, NLO and NNLO for the $pp\to H$ via gluon fusion process at $\sqrt{S} = 8$~TeV. Predictions of the Mellin-\chbar\ model with $\lambda_h=0.6$, 
compared to those of the standard \chbar\ method and of the scale-variation approach.}
  \label{fig:ggHMellin}
\end{figure}

This process is characterised by the dominance at every perturbative order of the gluon-gluon channel. The other partonic channels, 
which only enter at NLO and at NNLO order, turn out to only give subleading contributions. 
We calculate the dominant moment from the Higgs coefficient functions in Mellin space given in \cite{Kovacikova:2011gh} at $N_0=2$, this value giving the best approximation of the full Mellin inversion integral, as established using a saddle-point approximation~\cite{Bonvini:2010tp,Bonvini:2012an}.
We then compare the MHOUs thus obtained with those from the convoluted coefficients method in Table~\ref{tab:gghiggsMellin}. We see that the behaviour of the dominant Mellin moments at the various perturbative  orders is very similar 
to the one of the coefficients extracted after the convolution with the parton distribution functions. Indeed, as shown also in Figure~\ref{fig:ggHMellin}, the resulting uncertainty intervals 
in the two approaches agree very well.

\subsection{The Drell-Yan process at the LHC}

The Drell-Yan process for $Z$ production at the LHC is dominated by the $q\bar{q}$ channel both at LO (where it is the only channel) and 
at NLO, where also the quark-gluon channel starts to contribute.  
At NNLO also the gluon-gluon channel opens up.
It is known that the total net effect of the new channels on the total NNLO contribution is PDF-dependent, due to the uncertainties 
of the gluon PDFs. In particular, the NNLO contribution can change sign according to the PDF set used. 

\begin{figure} 
    \centering
    \includegraphics[width=0.7\textwidth]{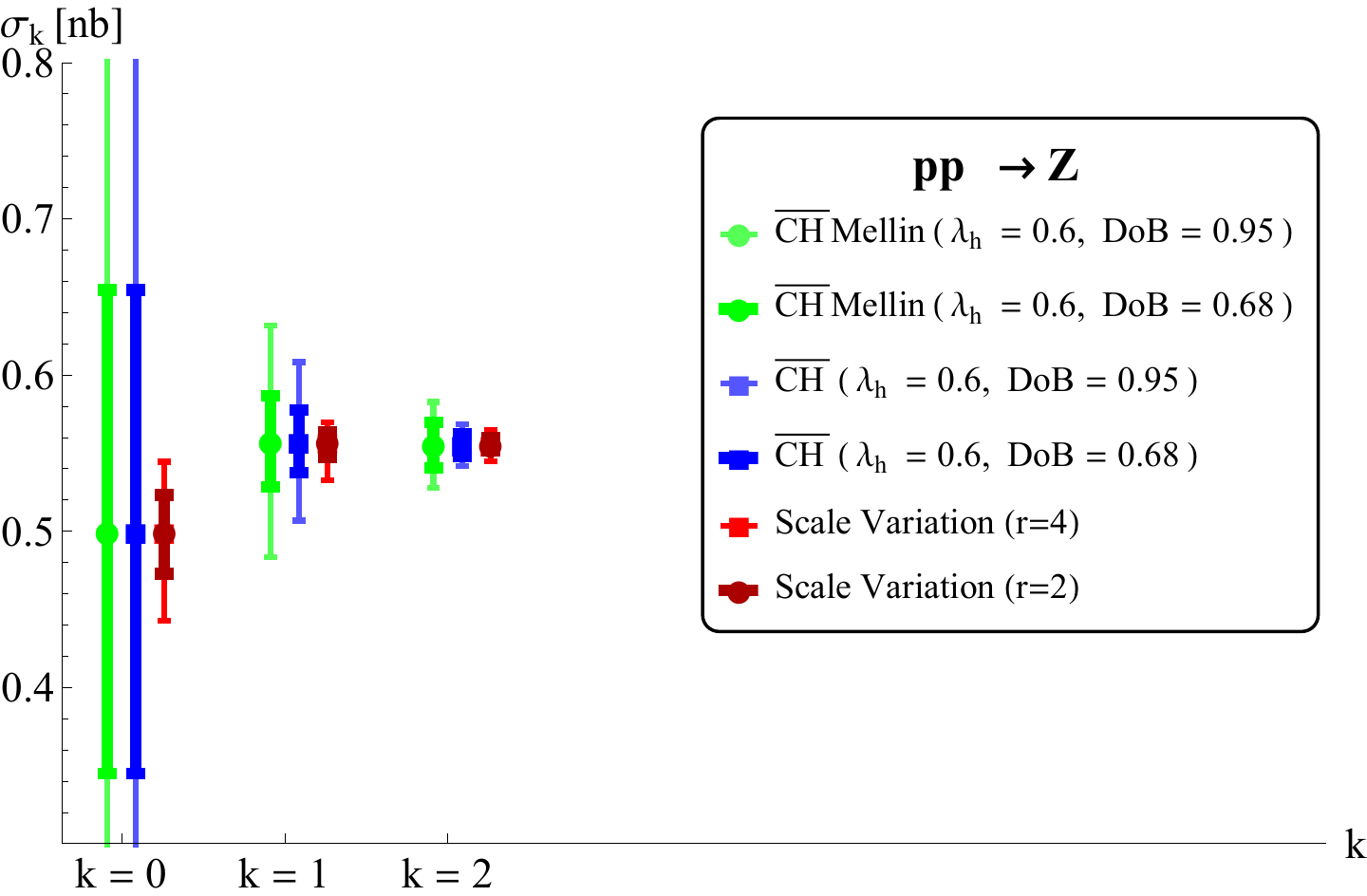}
	\caption{Size of the MHO uncertainty intervals at LO, NLO and NNLO for the $pp\to Z$ process at $\sqrt{S} = 8$~TeV. Predictions of the Mellin-\chbar\ model with $\lambda_h=0.6$, 
compared to those of the standard \chbar\ method and of the scale-variation approach.}
   \label{fig:DYMellin}
\end{figure}

In our case, using the NNLO 
NNPDF 2.3 set~\cite{Ball:2012cx}, it is negative due to the predominance of the negative quark-gluon channel at this order.
On the other hand, the Mellin-space coefficient function for the $q\bar{q}$ channel, which we use in the analysis, 
is always positive. Hence, it is not able to capture, alone,  the complex pattern of the perturbative expansion for this process. Indeed we see in Table~\ref{tab:DYMellin} and in
Figure~\ref{fig:DYMellin} that, 
starting from NLO, the uncertainty interval obtained with the Mellin method is systematically larger than the one obtained with the 
standard coefficient-based one.

One could in principle  work around this issue performing a Mellin analysis for each channel separately. However, to recombine 
the different uncertainties in order to get a single band for the complete cross section would then require  the knowledge of the weight of each channel. This in turn requires the use of the PDFs to determine the corresponding parton fluxes, introducing 
a dependence on long-range physics into the Mellin moment method, and therefore spoiling its main advantage.

\begin{table}

\centering
  \caption{Results for the analysis of Higgs production in proton-proton collisions and of Drell-Yan production of a Z 
   boson at $\sqrt{S} = 8$~TeV. We quote the perturbative order $k$ at which the observable is calculated,  the central value for the theoretical prediction at that order, the value of 
   the coefficient function at the dominant Mellin moment $N_0$  and the MHOs uncertainty intervals computed using 
   the \chbar\ model at 68\% DoB and 95\% DoB with both the Mellin moment method and the coefficient-based 
   approach.}
  \label{tab:resMellin}
  
\subtable[MHOUs for Higgs production in proton-proton collisions via gluon fusion. The perturbative series of the observable 
at the order $k$ is defined as $\sigma_k(pp \rightarrow H) = \sum_{n=2}^k \alpha_s^n h_n$.]
{
  \centering
  \ra{1.1}
  \begin{tabular}{@{}lcccccc@{}}\toprule
    \multicolumn{6}{c}{\text{\LARGE $pp \rightarrow H$}} \\
    \toprule
    Order &  $\sigma_k$[pb]	& $C(N_0=2)$ &\chbar\ Mellin, 68\%	& \chbar\ Mellin, 95\%	& \chbar, 68\%	& \chbar, 95\%	\\ \midrule
    $k = 2$		& 5.6	& 1	& $\pm 3.35$	& $\pm 21.46$	& $\pm 3.35$	& $\pm 21.46$		\\
    $k = 3$		& 13.3	& 12.12	& $\pm 4.54$	& $\pm 11.48$	& $\pm 4.51$	& $\pm 11.42$		\\
    $k = 4$		& 18.38	& 71.19	& $\pm 3.70$	& $\pm 6.99$	& $\pm 3.52$	& $\pm 6.65$		\\
    \bottomrule
  \end{tabular}
  \label{tab:gghiggsMellin}
}

\subtable[MHOUs for $Z$ production in the Drell-Yan process. The perturbative series of the 
observable at the order $k$ is defined as $\sigma_k(pp \rightarrow Z \rightarrow e^+e^-) = \sum_{n=0}^k \alpha_s^n h_n$.]{
  \centering
  \ra{1.1}
  \begin{tabular}{@{}lcccccc@{}}\toprule
    \multicolumn{6}{c}{\text{\LARGE $pp \rightarrow Z $}} \\
    \toprule
    Order &  $\sigma_k$ & $C(N_0=2)$	& \chbar\ Mellin, 68\%	& \chbar\ Mellin, 95\%	& \chbar, 68\%	& \chbar, 95\%	\\ \midrule
    $k = 0$		& 0.499	& 1	& $\pm 0.155$	& $\pm 0.991$	& $\pm 0.155$		& $\pm 0.991$		\\
    $k = 1$		& 0.557	& 2.92	& $\pm 0.029$	& $\pm 0.074$	& $\pm 0.020$		& $\pm 0.051$		\\
    $k = 2$		& 0.555	& 7.53	& $\pm 0.014$	& $\pm 0.027$	& $\pm 0.007$		& $\pm 0.013$		\\
    \bottomrule
  \end{tabular}
  \label{tab:DYMellin}
}

\end{table}

\clearpage

\clearpage

\section{Statistical uncertainty in the determination of $\lambda$}
\label{section:statuncapp}

In this Appendix we  detail the procedure through which we determine the uncertainty on our determination of the rescaling factor $\lambda$ (or $\lambda_h$) of the expansion parameter in the \chbar\ approach.
We recall that the
optimal value for  $\lambda$ is determined by comparing, for the \chbar\ model with a given value of $\lambda$, its measured success rate in describing the missing higher order uncertainty (i.e. in returning an interval that includes the known higher order result) with the value of the Degree of Belief given as an input. 

What we need to establish is the statistical uncertainty on the measured success rate, resulting from the finite size of the sample of observables that we employ in our survey.
For this purpose, we 
follow the procedure suggested in~\cite{Cameron:2010bh}. It relies on the  
assumption that the measured success rate, $s/n$, where $s$ is the number of successes and $n$ the size of the observables' sample, is a point estimator for $p$, the real proportion of successes in the underlying population.
The likelihood of observing 
a success rate value $s/n$, given the true value $p$, is then proportional to $p^s(1-p)^{n-s}$, and upon normalisation over
the interval $0<p<1$ it can be written as a Beta distribution, 
\begin{align}
   \label{eq:beta}
   B(s+1,n-s+1) & =\frac{(n+1)!}{s!(n-s)!}p^{s}(1-p)^{n-s}\, .
\end{align}
If we express our ignorance on the value of $p$ by means of a Bayes-Laplace uniform prior, $P_{pr}(p)=1$ 
over the interval $0<p<1$,  we can, upon application of Bayes' theorem, use  the normalised likelihood
function~(\ref{eq:beta}) as a posterior probability distribution. We can then define the lower and upper bounds, 
$p_l$ and $p_u$, of an equal-tailed  $c\%=1-\alpha$ credibility interval for $p$ through the relations
\begin{align} \int\limits_0^{p_l}&dp\,B(s+1,n-s+1)=\frac{\alpha}{2}\quad\text{and}\quad\int\limits_{p_u}^1dp\,B(s+1,n-s+1)=\frac{\alpha}{2}
\end{align}
respectively.

A proper application of this procedure to our problem would require the evaluation, for each value of $\lambda$ and for each measured $s/n$ rate, of the credibility interval $[p_l, p_u]$ for a given value of $\alpha$. In practice, in order to simplify both the calculation and the graphical representation of this uncertainty, we determine a 68.3\%-credible interval for an ideal curve where success rate = requested DoB. This is the interval that is represented as a grey band in the right hand plots of Figures~\ref{fig:CHApproxFacAll} and \ref{fig:globalhadronicNoDIS}. As long as the actual curves do not differ too much from this ideal one, one can easily gauge the size of the uncertainty, and therefore to what extent two curves obtained with two values of $\lambda$ are, or are not, significantly different.


\begin{thebibliography}{99}

\bibitem{Cacciari:2011ze}
  M.~Cacciari and N.~Houdeau,
  JHEP {\bf 1109} (2011) 039
  [arXiv:1105.5152 [hep-ph]].

\bibitem{Ball:2011us}
  R.~D.~Ball, V.~Bertone, L.~Del Debbio, S.~Forte, A.~Guffanti, J.~I.~Latorre, S.~Lionetti and J.~Rojo {\it et al.},
  Phys.\ Lett.\ B {\bf 707} (2012) 66
  [arXiv:1110.2483 [hep-ph]].

\bibitem{Goria:2011wa}
  S.~Goria, G.~Passarino and D.~Rosco,
  Nucl.\ Phys.\ B {\bf 864} (2012) 530
  [arXiv:1112.5517 [hep-ph]].

\bibitem{Forte:2013mda}
  S.~Forte, A.~Isgr\`o and G.~Vita,
  Phys.\ Lett.\ B {\bf 731} (2014) 136
  [arXiv:1312.6688 [hep-ph]].

\bibitem{David:2013gaa}
  A.~David and G.~Passarino,
  Phys.\ Lett.\ B {\bf 726} (2013) 266
  [arXiv:1307.1843].

\bibitem{Agashe:2014kda}
  K.~A.~Olive {\it et al.}  [Particle Data Group Collaboration],
  Chin.\ Phys.\ C {\bf 38} (2014) 090001.

\bibitem{Cacciari:2003fi}
  M.~Cacciari, S.~Frixione, M.~L.~Mangano, P.~Nason and G.~Ridolfi,
  JHEP {\bf 0404} (2004) 068
  [hep-ph/0303085].

\bibitem{Anastasiou:2005qj}
  C.~Anastasiou, K.~Melnikov and F.~Petriello,
  Nucl.\ Phys.\ B {\bf 724} (2005) 197
  [hep-ph/0501130].

\bibitem{Anastasiou:2002yz}
  C.~Anastasiou and K.~Melnikov,
  Nucl.\ Phys.\ B {\bf 646} (2002) 220
  [hep-ph/0207004].

\bibitem{PhysRevLett.73.1207}
  A.~I.~Vainshtein and V.~I.~Zakharov,
  Phys.\ Rev.\ Lett.\  {\bf 73} (1994) 1207
   [Erratum-ibid.\  {\bf 75} (1995) 3588]
  [hep-ph/9404248].

\bibitem{Zakharov:1992bx}
  V.~I.~Zakharov,
  Nucl.\ Phys.\ B {\bf 385} (1992) 452.

\bibitem{Fischer:1997bs}
  J.~Fischer,
  Int.\ J.\ Mod.\ Phys.\ A {\bf 12} (1997) 3625
  [hep-ph/9704351].

\bibitem{Beneke:1998ui}
  M.~Beneke,
  Phys.\ Rept.\  {\bf 317} (1999) 1
  [hep-ph/9807443].

\bibitem{Bonvini:2010tp}
  M.~Bonvini, S.~Forte and G.~Ridolfi,
  Nucl.\ Phys.\ B {\bf 847} (2011) 93
  [arXiv:1009.5691 [hep-ph]].

\bibitem{Bonvini:2012an}
  M.~Bonvini, S.~Forte and G.~Ridolfi,
  Phys.\ Rev.\ Lett.\  {\bf 109} (2012) 102002
  [arXiv:1204.5473 [hep-ph]].

\bibitem{Baikov:2008jh}
  P.~A.~Baikov, K.~G.~Chetyrkin and J.~H.~Kuhn,
  Phys.\ Rev.\ Lett.\  {\bf 101} (2008) 012002
  [arXiv:0801.1821 [hep-ph]].

\bibitem{Larin:1990zw}
  S.~A.~Larin, F.~V.~Tkachov and J.~A.~M.~Vermaseren,
  Phys.\ Rev.\ Lett.\  {\bf 66} (1991) 862.

\bibitem{Larin:1991tj}
  S.~A.~Larin and J.~A.~M.~Vermaseren,
  Phys.\ Lett.\ B {\bf 259} (1991) 345.

\bibitem{Biswas:2009rb}
  S.~Biswas and K.~Melnikov,
  JHEP {\bf 1002} (2010) 089
  [arXiv:0911.4142 [hep-ph]].

\bibitem{Baikov:2012er}
  P.~A.~Baikov, K.~G.~Chetyrkin, J.~H.~Kuhn and J.~Rittinger,
  Phys.\ Rev.\ Lett.\  {\bf 108} (2012) 222003
  [arXiv:1201.5804 [hep-ph]].

\bibitem{Chetyrkin:1994js}
  K.~G.~Chetyrkin, J.~H.~Kuhn and A.~Kwiatkowski,
  In *Geneva 1994, Reports of the working group on precision calculations for the Z resonance* 175-263
  [hep-ph/9503396].

\bibitem{Weinzierl:2009yz}
  S.~Weinzierl,
  Phys.\ Rev.\ D {\bf 80} (2009) 094018
  [arXiv:0909.5056 [hep-ph]].

\bibitem{Larin:1996wd}
  S.~A.~Larin, P.~Nogueira, T.~van Ritbergen and J.~A.~M.~Vermaseren,
  Nucl.\ Phys.\ B {\bf 492} (1997) 338
  [hep-ph/9605317].

\bibitem{Baikov:2005rw}
  P.~A.~Baikov, K.~G.~Chetyrkin and J.~H.~Kuhn,
  Phys.\ Rev.\ Lett.\  {\bf 96} (2006) 012003
  [hep-ph/0511063].

\bibitem{Baikov:2006ch}
  P.~A.~Baikov and K.~G.~Chetyrkin,
  Phys.\ Rev.\ Lett.\  {\bf 97} (2006) 061803
  [hep-ph/0604194].

\bibitem{Maierhofer:2012vv}
  P.~Maierh\"ofer and P.~Marquard,
  Phys.\ Lett.\ B {\bf 721} (2013) 131
  [arXiv:1212.6233 [hep-ph]].

\bibitem{Spira:1995rr}
  M.~Spira, A.~Djouadi, D.~Graudenz and P.~M.~Zerwas,
  Nucl.\ Phys.\ B {\bf 453} (1995) 17
  [hep-ph/9504378].

\bibitem{Spira:1995mt}
  M.~Spira,
  hep-ph/9510347.

\bibitem{Harlander:2003ai}
  R.~V.~Harlander and W.~B.~Kilgore,
  Phys.\ Rev.\ D {\bf 68} (2003) 013001
  [hep-ph/0304035].

\bibitem{Czakon:2013goa}
  M.~Czakon, P.~Fiedler and A.~Mitov,
  Phys.\ Rev.\ Lett.\  {\bf 110} (2013) 25,  252004
  [arXiv:1303.6254 [hep-ph]].

\bibitem{Catani:2009sm}
  S.~Catani, L.~Cieri, G.~Ferrera, D.~de Florian and M.~Grazzini,
  Phys.\ Rev.\ Lett.\  {\bf 103} (2009) 082001
  [arXiv:0903.2120 [hep-ph]].

\bibitem{Brein:2003wg}
  O.~Brein, A.~Djouadi and R.~Harlander,
  Phys.\ Lett.\ B {\bf 579} (2004) 149
  [hep-ph/0307206].

\bibitem{Campbell:1999ah}
  J.~M.~Campbell and R.~K.~Ellis,
  Phys.\ Rev.\ D {\bf 60} (1999) 113006
  [hep-ph/9905386].

\bibitem{Campbell:2002tg}
  J.~M.~Campbell and R.~K.~Ellis,
  Phys.\ Rev.\ D {\bf 65} (2002) 113007
  [hep-ph/0202176].

\bibitem{Ball:2011uy} R.~D.~Ball {\it et al.} [NNPDF Collaboration], 
  Nucl.\ Phys.\ B {\bf 855} (2012) 153
  [arXiv:1107.2652 [hep-ph]].

\bibitem{Ball:2012cx}
  R.~D.~Ball, V.~Bertone, S.~Carrazza, C.~S.~Deans, L.~Del Debbio, S.~Forte, A.~Guffanti and N.~P.~Hartland {\it et al.},
  Nucl.\ Phys.\ B {\bf 867} (2013) 244
  [arXiv:1207.1303 [hep-ph]].

\bibitem{Cacciari:2008gp}
  M.~Cacciari, G.~P.~Salam and G.~Soyez,
  JHEP {\bf 0804} (2008) 063
  [arXiv:0802.1189 [hep-ph]].

\bibitem{Kovacikova:2011gh}
  P.~Kovacikova,
  Fortsch.\ Phys.\  {\bf 59} (2011) 1070
  [arXiv:1104.4968 [hep-ph]].

\bibitem{Cameron:2010bh}
  E.~Cameron, 
Publications of the Astronomical Society of Australia, 28 (2011) 128
  [arXiv:1012.0566 [astro-ph.IM]].

  
\end{thebibliography}
\end{document}